\documentclass{nature}

\usepackage{amsmath,amssymb}
\usepackage{epsfig}
\usepackage{epstopdf}
\usepackage{graphicx}
\usepackage{dcolumn}
\usepackage{bm}
\usepackage{float}
\usepackage{caption}
\usepackage{subcaption}
\usepackage[plainpages=false,pdfpagelabels,colorlinks=true,linkcolor=blue,urlcolor=blue,citecolor=blue,pdftitle={Title},pdfauthor={},pdfdisplaydoctitle=true,pdfduplex=DuplexFlipLongEdge]{hyperref}
\usepackage{adjustbox} 
\usepackage{rotating}  
\usepackage{grffile}	
\usepackage{mathtools,units}
\usepackage{tabularx}

\usepackage[left,pagewise,displaymath,mathlines]{lineno}

\usepackage{soul}   

\newcommand{\onlinecite}[1]{\hspace{-1 ex} [\nocite{#1}\citenum{#1}]}

\newcommand{\trt}{TaRhTe$_4$}

\newcommand{\tit}{TaIrTe$_4$}
\newcommand{\nit}{NbIrTe$_4$}
\newcommand{\nrt}{NbRhTe$_4$}

\title{Fermi energy Weyl nodes in $\mathbf{AM}$Te$_4$ ($\mathbf{A}$=Ta, Nb, $\mathbf{M}$=Ir, Rh)}

\author{Shivam Parasar$^{1}$, Jeroen van den Brink$^{2,3,*}$ \& Rajyavardhan Ray$^{1,2,\ddagger}$}

\makeatletter
\let\saved@includegraphics\includegraphics
\AtBeginDocument{\let\includegraphics\saved@includegraphics}
\renewenvironment*{figure}{\@float{figure}}{\end@float}
\makeatother

\begin{document}

\maketitle

\begin{affiliations}
\item{Department of Physics, Birla Institute of Technology Mesra, Ranchi, India - 835215}
\item{Leibniz Institute for Solid-State and Materials Research (IFW) Dresden, Helmholtzstr. 20, 01069 Dresden, Germany}
\item{Institute of Theoretical Physics and W{\"u}rzburg-Dresden  Cluster of Excellence {\it ct.qmat}, Technische Universit{\"a}t Dresden, 01062 Dresden, Germany} 
\end{affiliations}

\noindent
Email: j.van.den.brink@ifw-dresden.de, r.ray@bitmesra.ac.in

\noindent
\date{\today}

\noindent



\begin{abstract}
Key aspects of the quantum oscillations and magnetoresistance in Weyl semimetals $\mathbf{AM}$Te$_4$ ($\mathbf{A}$=Nb,Ta, $\mathbf{M}$=Rh, Ir) persist as open questions, 
obscuring the link between their topological electronic structure and practical implementations.
Employing a generalised search procedure, we carry out a comprehensive scan 
of WPs accounting for all the subbands close to the Fermi energy, and show 
that this dramatically alters the WP landscape in these compounds. In particular, we predict
these compounds to feature WPs within a few meV of the Fermi energy which  
significantly influence their properties. 
Remarkably, most of the considered compounds host WPs of more than one type, including {\nit} which hosts type-I, II and III Weyl points.
Our comparative analysis of structure and fidelity of computational parameters/models not only provides a detailed mapping of the complex 
electronic structure in these compounds, but also clarifies quantum oscillations and magnetoresistance observations in this family, 
bridging the gap between theory and experiments and offering a framework for precise tunability of WPs.
\end{abstract}


\section*{Introduction}
Weyl semimetals are a class of topological quantum materials featuring linear band crossings, called Weyl 
points (WPs).\cite{Herring1937, Yan2017, Armitage2018} WPs act as sources of Berry curvature, carry topological charge $\chi=\pm 1$, display spin-momentum 
locking on the so-called Weyl cones around the points, and are robust against small perturbations. They can 
be created/annihilated in pairs such that the net chirality over the Brillouin zone (BZ) remains zero.\cite{Nielsen1981} Further,
 the Fermi arcs --- projected surface states connecting the WPs are also topological, and depending on their titling features may feature a 
 diverging Berry curvature.\cite{Wawrzik2021} As a result, WSMs are considered promising for 
electronic, photonic and spintronic applications, as well as from a fundamental viewpoint.\cite{Zhong2025}

WPs can be categorized into different categories depending on the tilt of the cone and curvature of the participating bands. 
While Type-I WPs posses point-like Fermi surface(s),\cite{Wan2011} type-II WPs are formed at the intersection of 
electron-hole pockets due to tilting of the Weyl cones.\cite{Soluyanov2015} At a critical tilt between type-I and type-II,
band curvature effects can lead to intersections between electron-electron or hole-hole fermion pockets, forming the so-called 
type-III WPs.\cite{Li2021, Li2022}
 
{\tit} has long been considered a paradigmatic type-II Weyl semimetal hosting the minimal number of symmetry-allowed 
WPs\cite{Koepernik2016} and large observed Fermi arcs.\cite{Haubold2017} 
{\tit} exhibits unconventional surface superconductivity and superconducting Fermi arcs where Cooper pairing occurs within the 
topological surface states.\cite{Xing2019, Kuibarov2024} This topological state is quite robust, as evidenced by the nearly 
identical superconducting transitions observed in both {\tit} and its sister compound {\nit} ($A$IrTe$_4$) when tuned under 
high pressure.\cite{Long2021, Mu2021} {\tit} further exhibits large photoresponsivity,\cite{Ma2019} a large out-of-plane 
damping-like spin-orbit torque at room temperature,\cite{Bainsla2024} and spin charge conversion\cite{Tang2025} or 
magnetoresistive switching\cite{Li2024} in heterostructures. 
Beyond equilibrium transport, the topological properties can be 
exploited to achieve room temperature nonlinear Hall effect\cite{Kumar2021} as well as high-sensitivity nonlinear responses and 
room-temperature terahertz sensing.\cite{Jiang2025} 

With regard to the electronic structure, a good qualitative and quantitative agreement between density functional theory (DFT) calculations using the local 
density approximation (LDA) and angle-resolved photoelectron spectroscopy (ARPES) has been established in {\tit} 
considering charge (electron count) conserving shifts of bands up to 75 meV.\cite{Haubold2017, Belopolski2017} A remarkable 
aspect is the observation of Fermi arcs in proximity to their predicted location. 

Comparison with quantum oscillation measurements further corroborates the energy shifts required for a reasonable agreement, 
implying that the WP quartet predicted at $82.7$ meV may actually be present only at $40$ to $50$ meV above 
the Fermi energy. However, the angular dependence of the de Haas-van Alphen (dHvA) oscillations reveals features 
that can not be satisfactorily explained by the available DFT based calculations.\cite{Khim2016} 
Specifically, some of the observed frequencies, including the lowest frequency, $F_0 = 10.8$ T, corresponding to the quantum limit were
not captured by the DFT calculations.
At the same time, magnetoresistance (MR) data shows that $\rho_{} \propto B^{1.5}$, deviating from the ideal $B^2$ dependence
expected for a compensated semimetal with paired electron-hole pockets near the Fermi energy.\cite{Pletikosic2014,Zhang2019}

Its isostructural and isoelectronic cousin {\trt}, on the other hand, is predicted to host three sets of WP 
quartets, lying at $-43$ meV, $-63.7$ meV, and $125.2$ meV.\cite{Zhang2024} MR 
measurements reveal a characteristic $B^{2}$ dpendence of 
cyclotronic origin.\cite{Son2013, Behnami2025} Moreover, the observed large negative longitudinal MR for different directions of parallel electric 
and magnetic fields is consistent with chiral anomaly. Interestingly, similar to {\tit}, the Nb counterpart, {\nit}, also shows positive MR.
However, it scales as $B^{1.25}$ and the values are much smaller than in {\tit}.  

The fact that signals from the electron pockets ($E_4$ and $E_5$) in the calculated electronic structure of {\tit} 
are absent in dHvA measurements, while there is no trace of some 
observed frequencies ($F_1$ and $F_2$) in the calculations,\cite{Khim2016} raises fundamental concerns regarding its electronic structure. 
Further,  how does one reconcile the fact that 
the Landau level (LL) fan diagram yields a phase intercept of $\gamma \approx 0$, which is unexpected for a Type-II Weyl 
semimetal hosting Weyl Points (WPs) within the $40$ to $50$ meV range. This apparently trivial phase contradicts the system’s extreme 
transport signatures: an abrupt $\sim 11$ T Hall kink (the Quantum Limit) and a non-saturating $B^{1.5}$ MR 
persisting to 70 T. The origin of the MR plateau at $50$ to $60$ T also remains unidentified.

This MR response differs from the {\trt} homologue, which displays a clean $B^2$ dependence and a robust chiral anomaly. 
Can these difference be uniquely attributed to the energy distribution of the multiple Weyl quartets predicted in {\trt}? 
Concurrently, while {\nit} exhibits large positive MR similar to {\tit},
balanced charge carriers as a possible origin has been ruled out.\cite{Zhou2019} It is notable that despite the fact that the energy positions of the predicted WPs 
in both the Ir-based compounds is similar, MR in 
{\nit} is about 10 times smaller than in {\tit}. Further, a non-zero phase intercept ($\gamma \ne 0$) in 
Shubnikov-de Haas (SdH) oscillations is observed which confirms its topological electronic structure, in sharp contrast with the observations in {\tit}. 
These fundamental inconsistencies highlight critical issues with the 
current mapping of the topological landscape for these semimetals, which further bottlenecks efforts to harness 
their topological properties. Therefore, for potential applications of these materials --- the AMTe$_4$ 
family --- a clear understanding of the electronic structure and possible tunability across 
the members are highly desirable.

Very recently, based on DFT calculations using generalized gradient approximation (GGA) with
$22 \times 6 \times 6$ $k$-mesh, 16 WPs have been predicted for {\tit}.\cite{Pandey2025} Two quartets lie in the $k_z=0$ plane, 
at energies $-45.78$ meV and $103.09$ meV, and one octet ($k_z \ne 0$) lies at $-46.09$ meV. The discrepancy between 
the initially reported 4 and their predicted 16 WPs, were attributed to the quality of Wannier models. 

The commonality between the above density functional studies is that they focus only on the band crossings between 
the topmost valence band $N$ and the lowermost conduction band $N+1$. In fact, this particular aspect is shared by 
(almost) all the DFT-based studies of Weyl semimetals in general. In ideal cases where the Fermi surface is formed only by bands 
$N$ and $N+1$, it is expected that their crossings lie close to the Fermi energy and contribute dominantly to response 
functions. However, real materials are complex and often deviate from ideal situations. We will show indeed that in the $AM$Te$_4$ 
systems, $N-2$ and $N-1$ bands must be accounted for as well.

Here, using the ternary Weyl semimetals, $AM$Te$_4$ ($A$=Ta, Nb, $M$=Ir, Rh), as an example, we show that 
crossings between sub-bands may occur close to the Fermi energy. 
Remarkably, in three out of the four considered compounds, we find stable WPs merely $\lesssim \pm 17$ meV from the Fermi energy, some lying even
$-1.5$ meV (about 17K) away.
 In all the compounds, many more WPs are found within the investigated energy range, completely altering the known WP landscape.

In view of these findings, we revisit the dHvA calculations in \tit\ using our Wannier model(s) and find a significantly improved 
qualitative and quantitative agreement between theory and experiment(s). 
Our comparative strategy to investigate the effects of the choice of functionals (LDA or GGA), crystal structure,
and the quality of the employed Wannier models on electronic properties of AMTe$_4$ further provides
insights into how WPs transform, and their energy and position uncertainties in the density functional calculations, 
facilitating their identification and/or signatures in the experiments along with a clear understanding of the material's responses.

\section*{Crystal structures and Symmetries}

$AM$Te$_4$ crystallizes in a layered, noncentrosymmetric orthorhombic crystal structure, with $Pmn2_1 (31)$ 
space group and is shown in Fig. 1(a). The structure contains four 
formula units per unit cell which forms two layers along the c-axis. Within each layer, the transition metal ions are octahedrally coordinated by Te, leading
to a network of distorted edge-sharing $TM$-Te ($TM$: transition metals) octahedra.

The crystal structure lacks inversion, but possesses a mirror symmetry $m(x)$ in the $y-z$ plane, a glide plane 
formed by mirror symmetry $m(y)$ in the $x-z$ plane followed by a translation by (1/2, 0, 1/2). There also exists 
a non-symmorphic (screw) rotation symmetry, $C_2(z)$ + (1/2, 0, 1/2). These symmetries, together with time-reversal symmetry, leads to a 
minimal possible quartet of WPs (degeneracy of four) lying in a plane.\cite{Soluyanov2015, Koepernik2016} 
In general, however, a WP at an arbitrary ($k_x, k_y, k_z$) should be 8-fold degenerate.

\section*{Bulk electronic structure}
In order to disentangle the roles of crystal structure and choice of DFT parameters, such as $k$-mesh and the exchange-correlation (XC) functionals, we systematically 
investigate and compare the bulk electronic properties. All the density functional calculations were done using the Full-Potential Local-Orbital (FPLO) 
code \cite{Koepernik1999} [\href{https://www.fplo.de}{https://www.fplo.de}] (see Methods for details). Fig. 1(b) shows the electronic bandstructure of \tit$^*$, calculated using the 
experimental crystal structure (indicated by $*$ here and after). The results were obtained via DFT using local density approximation (LDA)\cite{PW1992} and a $k$-mesh with $20\times10\times10$
 intervals. The Fermi 
surface consists of nested electron and hole pockets along $\Gamma-X$, small hole pockets along $S-Y$ and $Y-\Gamma$. The fish-and-toad hole pocket along 
$Y-\Gamma$ almost spans the entire BZ length. 

A comparison with a $12\times 6 \times 6 $ $k$-mesh,\cite{Koepernik2016} is shown in Fig. 1(d) and Fig. S1.
We observe that the Fermi surface (FS) topology is sensitive to the $k$-mesh density.
Specifically, employing a finer mesh results in a smaller electron pocket and a larger hole pocket along $\Gamma-X$  with an upward shift of  $\sim 15$ meV
concurrent with enlarged pocket along $S-Y$ with an upward shift of $\sim 10$ meV. Notably, the directions of these shifts due to the choice of the $k$-mesh are in agreement with ARPES 
observations, albeit somewhat smaller than desired.\cite{Haubold2017} Taking a finer mesh also introduces the fish-and-toad hole pocket spanning almost 
the entire $Y-\Gamma$ length, which was absent in the case of a coarse mesh (see Fig. S1). 

\begin{figure}[hb!] 
    \centering
    \includegraphics[ width=0.995\textwidth]{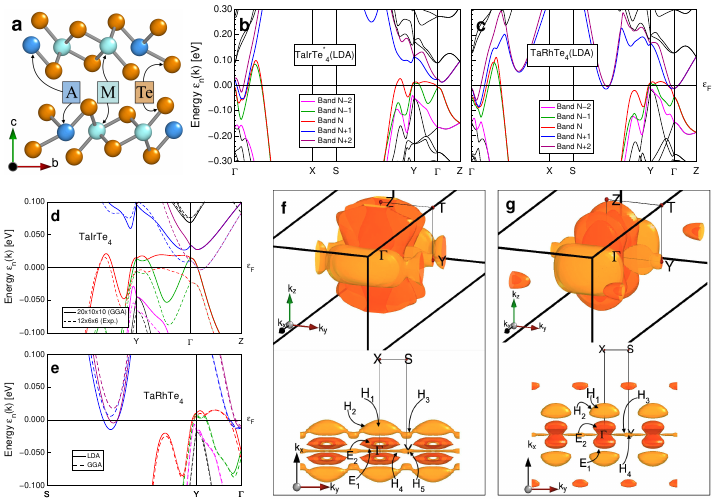}
	\caption{\small {\textbf{Crystal structure and electronic properties.} \textbf{(a)} Crystal structure of $AM$Te$_4$ ($A$=Ta, Nb, $M$=Ir, Rh). \textbf{(b-c)} Electronic
	band structure for (b) the experimental crystal structure of {\tit}, and (c) the optimzed crystal structure of {\trt}, using LDA and $20\times 10 \times 10$ $k$-mesh. 
	\textbf{(d)} Band structures of {\tit} for the optimized crystal structure with GGA and $20\times10\times10$ $k$-mesh (solid line), and the experiemntal crystal structure with LDA and $12\times6\times6$ $k$-mesh (dashed line). 
	\textbf{(e)} Comparison between 
	band structure of {\trt} using LDA (solid line) and GGA (dashed line) exchange correlation functional for  $20 \times 10 \times10$ $k$-mesh. 
	\textbf{(f)} Calculated Fermi surface of {\tit} within GGA. \textbf{(g)} Calculated Fermi surface of {\trt} within LDA. 
	}}
\label{fig:1}
\end{figure}

For the optimized structure where only the atomic positions were optimized, this hole pocket is further stretched to span the entire $Y-\Gamma$ 
length, leading to open FS across the $k_y=\pm\pi$ BZ boundary, (see Supplementary Note II for details\cite{esi}). Additionally, 
there is an updward shift of $\lesssim 15$ meV along $\Gamma-X$ path in BZ in the relaxed structure. Switching to GGA introduces further 
quantitative changes $\lesssim 10$ meV. As shown in Figure \ref{fig:1}(d) and Fig. S1, accounting for BZ sampling, optimized structure, and 
the choice of XC-functional changes the Fermi surface topology from the earlier reports.\cite{Koepernik2016} 

Summarizing, the FS topology turns out to be quite sensitive to the choice of $k$-mesh while the choice of XC-functional brings about marginal quantitative changes.
The combined effect of employing the GGA functional with a denser $k$-mesh on the optimized structures leads to an upward shift of  
$\sim 25$ meV and $\sim 10$ meV, respectively, for the electron and hole pockets along $\Gamma-X$. At the same time, the hole pocket along $S-Y$ are shifted 
up by about $\sim 5$ meV, and a fish-and-toad like hole pocket spanning across the entire $Y-\Gamma$ BZ length appears.

The Fermi surface for the optimized structure of {\tit}, obtained using GGA on a $k$-mesh with $20 \times 10 \times 10$ intervals, is 
shown in Fig. \ref{fig:1}(f). The large pockets along 
$\Gamma-X$  and the donut-like feature near the $\Gamma$ point corresponds to the nested hole and electron pockets seen in the band structure. The sharp 
needle like feature arises from the hole-pocket along $Y-\Gamma$. Notably, the hole pocket along $S-Y$ 
is touching (merging with) the bigger hole pocket. 

In comparison, Fig. \ref{fig:1}(c) and Fig. \ref{fig:1}(g), respectively, show the electronic band structure and the Fermi surface of \trt, calculated using 
LDA and a $k$-mesh with $20 \times 10 \times 10$ intervals for the optimized crystal structure (see Methods for details).
An overall qualitative similarity with {\tit} is noted: The Fermi surface consists of nested electron
and hole pockets along $\Gamma-X$, small nested electron pockets along $S-Y$ and a small hole pocket with needle like dispersion along $Y-\Gamma$.
A distinct feature is the overall lowering of Band $N+1$ in {\trt}, forming electron pockets at $\Gamma$ and along $S-Y$. The former has a dumbell-like 
shape as oppose to the donut-like shape in {\tit}. 
A comparison between LDA and GGA is shown in Fig. \ref{fig:1}(e) and Fig. S2. The primary difference due to the choice of XC-functional
is a slight upward shift of band $N+1$ such that the small electron pocket along $S-Y$ disappears.

\section*{Weyl point landscape in Ta-based compounds}
We now turn our attention to the primary focus of this study --- exploration of WP landscape.
While Ta-based compounds have been studied extensively via DFT over the years,\cite{Koepernik2016, Khim2016, Haubold2017, Liu2016, Zhang2024, Pandey2025} investigative efforts have 
been exclusively focused on Weyl nodes originating from the topmost valence band $N$, {\it i.e.}, crossings between the topmost valence band
$N$ and the lowermost conduction band $N+1$, where $N$ represents the total number of electrons in the system. 
However, as evident from Fig. \ref{fig:1}(b) \& (c), bands $N-2$, $N -1$, and $N+1$ also lie within $\sim \pm 100$ meV around the Fermi 
energy, and may host WPs as well. Therefore, a comprehensive scan of WPs in these compounds must also include such band crossings.

\begin{figure}
\centering
    \includegraphics[width=0.96\textwidth]{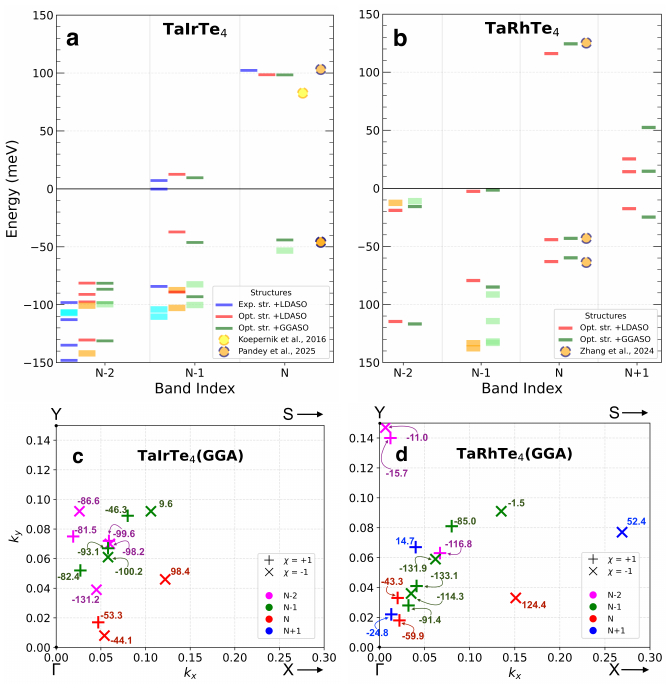}	
	\caption{\small \textbf{WP landscape for Ta systems.} \textbf{(a-b)} Energy distribution of WPs for (a) {\tit} and (b) {\trt}, accounting for different crystal structures and exchange-correlation functional. WP quartets are represented by dark thin bars while octets are represented by light-colored thick bars. \textbf{(c-d)} The in-plane coordinates ($k_x,k_y$) of the WPs in (c) {\tit}, and (d) {\trt}, obtained using GGA.} 
    \label{fig:wp_ta}
\end{figure}

Indeed, as shown in Fig. \ref{fig:wp_ta}, accounting for all the subbands lead to a large number of WPs in these systems. 
Remarkably, some of these WPs lie within a few meV around the Fermi energy. This is one of the primary findings of this 
study. In order to ascertain their 
robustness, we analyze in detail the influence of the crystal structure (atomic positions) and XC-functional on the WP energy 
and their positions in the BZ. 
For brevity, all the calculations correspond to $k$-mesh with $20 \times 10 \times 10$ intervals in the following unless mentioned otherwise. 

\subsection{\tit . }
Figure \ref{fig:wp_ta}(a) shows all the predicted WPs in {\tit}, accounting not only for the sub-bands but also different structures and DFT 
approximations. For the experimental structures,\cite{Koepernik2016, Haubold2017, Khim2016} LDA leads to a quartet of WPs at $102.3$ meV
from band $N$, consistent with the earlier report of $82.7$ meV.\cite{Koepernik2016} The energy difference of $\sim 20$ meV arises due to the 
denser $k$-mesh employed here. The WP is robust against minor changes in atomic positions up to 0.04 {\AA} (see Supplementary Note I\cite{esi}): 
even for the optimized structures,
only a tiny shift of $\sim 4$ meV is found. In sharp contrast, GGA calculations find a total of 16 WPs between these bands --- two quartets at
$98.4$ and $-44.1$ meV and an octet at $-53.3$ meV, which aligns with the recent report.\cite{Pandey2025} It is apparent that
this discrepancy in the total number of WP from Band $N$ arises due to the choice of XC-functional rather than any other factor
as previously suggested.\cite{Pandey2025}

Extending our analysis to include sub-band crossings near the Fermi level, we note that WPs from band $N-1$ (crossings between bands $N-1$ and $N$)
 lie between $12.6$ meV and $-110.3$ meV. 
LDA calculations on the experimental structure predicts three quartets and two octets, leading to a total of 28 WPs. Out of these, two 
quartets lie extremely close to the Fermi energy, at $-0.2$ meV, and another at $+7.2$ meV. It is interesting to note that while the WP above the 
Fermi energy is quite robust against changes in atomic positions as well as the choice of XC-functional (change only by about 5meV), the 
Fermi energy WP at $-0.2$ meV in the experimental structure shifts to $-37.2$ (LDA) and $-46.3$ meV (GGA) in the relaxed structures, suggesting possible tunability 
under strain. In general, WP energies change at most by about $\pm 20$ meV between LDA and GGA. The nature of these Fermi energy WPs is, however, independent of 
the choice of the functional. 

Further, all the WPs arising from band $N-2$ (intersection of bands $N-2$ and $N-1$) lie below $-80$ meV. Similar to the WPs from band $N-1$, the WP 
energy is sensitive to the atomic positions, leading to a shift of approximately $+30$ meV. Both LDA and GGA for optimized structures are in close 
agreement with each other. 

The BZ position of these WPs are listed in Table S6, where a comparison is drawn between the experimental and optimized structures. Figure \ref{fig:wp_ta}(c) shows the in-plane coordinates of all the WPs in {\tit}, 
obtained using GGA for the optimized structures. While many WPs lie close to the axes and may be susceptible to perturbations due to proximity 
to their (opposite) chirality partners,\cite{Zhang2024}
the Fermi-energy WPs from band $N-1$, at $9.6$ meV and $-46.3$ meV, are remarkably well-separated. They are, thus, expected to be robust against perturbations. Due to the large recipriocal distance between their chirality partners, spanning $\gtrsim 1/3$ of BZ length, {\tit} also hosts large Fermi arcs 
due to these Fermi-energy WPs, comparable in recirpocal length to the observed one between WPs from band $N$.\cite{Haubold2017} 

Within GGA for the optimized structure, with a dense $k$-mesh, band $N$ hosts both type-I and type-II WPs as shown in Fig. S4. Importantly, the 
WP $W_1$ at $E_{W1} \sim 100$ meV is a type-I WP whereas the other two WPs below $E_{\rm F}$ are type-II. In earlier calculations, $W_1$ was characterized 
to be type-II.\cite{Koepernik2016} We trace the origin of this discrepency to the seemingly minor structural differences with the experimental crystal 
structure (see Supplementary Note I). As also shown in Fig. S4, $W_1$ remains a type-I WP even within the LDA description of the optimized crystal structure, 
while the experimental structure clearly shows type-II nature. 
This high sensitivity of the nature of $W_1$ to the atomic positions suggests immediate impact on the transport studies depending on the quality of the 
synthesized samples. In comparison, the Fermi energy WPs from band $N-1$, lying with $\pm 50$ meV ($|E_{WP}| < 50$ meV), are type II.

\subsection{\trt .} The Rh-compound, {\trt} provides a natural way to understand how WPs transform across the family. Fig. \ref{fig:wp_ta}(b) shows
WPs from all the considered bands in {\trt}. The corresponding positions of these WPs in the BZ are listed in Table S9.  
In {\trt}, GGA and LDA agree closely with each other with regard to the WP energy as well as their positions in the BZ, 
with the largest shift of $\sim 30$ meV due to the choice of functional.
It is interesting to note a qualitative similarity in the energy distribution of WPs from band $N$ obtained within GGA for both Ta-based systems with relaxed atomic positions.   

It is quite remarkable that band $N-1$ in {\trt} also hosts a Fermi energy WP, lying at  $-2.6$ meV (LDA), or at $-1.5$ meV (GGA). Further, 
the separation between the two WPs closest to Fermi energy has increased from $55.9$ meV to $83.5$ meV within GGA.
Compared to {\tit}, where LDA and GGA differ in the number of WPs from band $N$, here we find such discrepancy arising in the number of WPs from band $N-1$.

Band $N-2$ in {\trt} hosts WPs at approximately $\sim -10$ meV and at $\sim -120$ meV. 
At the same time, band $N+1$ hosts WPs in an energy window of $-24.8$ meV and $52.4$ meV in GGA (In LDA, the corresponding energy range is smaller). 
Presence of multiple WPs in close proximity to the Fermi energy may significantly contribute to the transport and optical responses in {\trt}.

As shown in Fig. \ref{fig:wp_ta}(d), many WPs lie close to their partners of opposite chirality and may annihilate under perturbations.\cite{Zhang2024} Some of these
WPs lie close to the Fermi energy, such as those from band $N-2$. However, similar to the situation for {\tit}, the Fermi-energy 
WPs from band $N-1$ are well-separated and should not only be robust but also presumably lead to large Fermi arcs. In addition, the WPs from 
band $N+1$ at $52.4$ meV are also remarkably well-separated.
The low energy dispersions of the WPs from band N show that the WPs lying above $E_{\rm F}$, $W_1$ at 116 meV, is type-I, while both the WPs below $E_{\rm F}$ 
($W_2$ and $W_3$ at $-44$ and $-63$ meV, respectively) are type-II --- similar to {\tit}. This implies that {\trt} is a hybrid Weyl semimetal hosting WPs of 
different type. On the other hand, Fermi energy WPs with $|E_{WP}| < 50$ meV from bands $N-2$, $N-1$ and $N+1$ are type-II in nature.

To summarize, we predict the existence of robust type-II WPs in {\tit} and {\trt} close to the Fermi energy due to sub-band crossings.
In {\tit}, the Fermi energy WPs from band $N$ and $N-1$, are more sensitive to atomic arrangements rather than their DFT description. 
Compared to these, WPs from band $N$ not only seem to be sensitive to the choice of functional in DFT calculations, the above Fermi energy WP turns out to
be type-I in the optimized structure. In total, we find 20 WPs in {\tit} (four quartets and one octet) between $\pm 70$ meV. Below $-70$ meV, a swarm of 
WPs exist, originating from bands $N-2$ and $N-1$. In comparison, {\trt} is also a hybrid Weyl semimetal, hosting both type-I and type-II WPs from band $N$. 
In general, WPs in {\trt} are well-separated in energy in both LDA and GGA. We find a total of 36 WPs between $\pm 70$ meV, out of which 24 lie 
within $\pm 30$ meV ($\sim 350$ K). 

Remarkably, the Fermi energy WPs from band $N-1$ are well-separated in both {\tit} and {\trt} and are expected to feature large Fermi arcs directly 
measurable in experiments. It is also interesting to note similarities in the energy and BZ positions of
some of the WPs between {\tit} and {\trt}, suggesting chemical pressure as a viable route to tune these WPs. While signatures of these WPs in transport 
and nonlinear optical responses should be observable, the obtained WP landscape suggests an intricate dependence. The relatively small reciprocal distance 
between many of the predicted WPs and their (opposite) chirality partners suggests sensitivity to perturbations. As a result, their contributions to the 
transport and optical responses may depend heavilty on the sample quality.

\section*{Discussions}

\subsection{Angular dependence of dHvA frequencies. }
Sizable shifts of bands, upto $25$ meV either due to the choice of $k$-mesh, or their density functional description (LDA/GGA), 
leads to changes in the sizes of the fermion pockets which can be directly accessed by quantum oscillation experiments. Therefore, to validate 
our DFT findings we calculate the dHvA frequencies as a function of 
magnetic field direction and compare with the available experimental results for {\tit}.\cite{Khim2016} 

\begin{figure}
    \centering
    \includegraphics[width=0.98\textwidth]{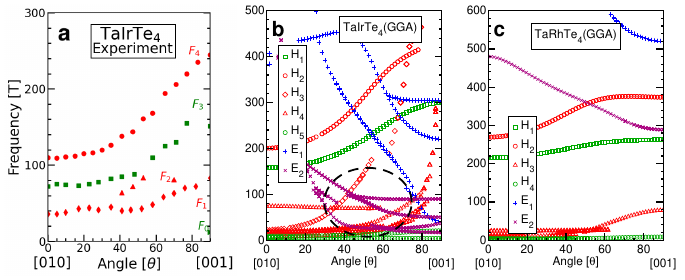}
	\caption{\small \textbf{Angular dependance of dHvA frequencies in Ta-based systems. } {\bf (a)} The observed dHvA frequencies from Ref. \onlinecite{Khim2016}. 
	\textbf{(b-c)} Calculated angular dependance of dHvA for (b) optimized structure of {\tit}, and (c) otpimized structure of {\trt}. In all cases, $k$-mesh with $20 \times 10 \times 10$ intervals were used 
	together with GGA.}
    \label{fig:dhva}
\end{figure}

The observed angular dependence of Fermi surface in {\tit} is shown in Fig. \ref{fig:dhva}(a), when the magnetic field ($B$) is rotated from 
$b$-axis ($\theta=0^{\circ}$) towards the $c$-axis ($\theta=90^\circ$) in the $bc$-plane.\cite{Khim2016}
Figures \ref{fig:dhva}(b) shows the corresponding results from DFT calculations using optimized structure and GGA with a dense $k$-mesh containing $20 \times 10 \times 10$ intervals.
The curves $F_3$ and $F_4$ in the experimental data were earlier ascribed to the nested hole pockets $H_1$ and $H_2$ along 
$\Gamma - X$, obtained using LDA and $12\times 6\times 6$ $k$-mesh.\cite{Khim2016} As explicitly shown in the Supplementary Note III\cite{esi}, the overall trends 
of $H_1$ and $H_2$ in Fig. \ref{fig:dhva}(b) is consistent with the earlier calculations. Quantitative differences are a direct results of small shifts 
of these hole pockets changing their areas by a small amount due to the finer $k$-mesh and GGA functional used in our calculations, 
as discussed earlier. For brevity, the fermion pockets have been explicitly labeled in Fig. \ref{fig:1}(f) \& (g).

The situation for other observed frequencies is somewhat complex due to the intricate details of the hole pockets formed by band $N$, and a clear correspondence in 
density functional calculations has remained challenging.\cite{Khim2016} Recall that, within GGA, the hole pocket $H_3$ lies along $S-Y$, the hole
pockets $H_4$ corresponds to the open needle-like pocket along $Y-\Gamma$ which also has a small but finite extend along $S-Y$, while the hole
pocket $H_5$ is small and centered at $Y$.
The resulting frequencies in the dHvA calculations take a wedge shape at intermediate angles between $35^{\circ}$ and $60^{\circ}$, and has 
been highlighted in Fig. \ref{fig:dhva}(b) with a dashed circle. This feature is remarkably similar to that of $F_1$ and $F_2$ in the experimental data. 
Similarly, one can also identify the $H_5$ curve in the calculations, due to the small hole pocket formed by band $N-1$ around the $Y$ point, with the 
observed $F_0$ provided that this pocket is shifted down such that no measurable signals can be generated except at small angles around $90^{\circ}$. 

An important distinction between the observed and calculated frequencies is the ``divergent" behavior of $H_2$, $H_3$ and $H_4$ at $\theta=90^{\circ}$ 
which arises due to their open FS topology. 
Therefore, for a quantitative comparison, 
the bands (and therefore the fermion pockets) needs to be shifted in a non-uniform manner. 
While the curve corresponding to $H_3$ needs to be shifted down, implying a smaller Fermi surface area and corresponding frequency, the hole pocket 
$H_4$ should not be open, much like features in the bandstrucuture calculations for the the experimental structure (Fig. S3). 
It is further evident that the required shifts are anisotropic.

The energy shifts required for a quantitative agreement between the observed and the calculated dHvA frequencies suggests, that the hole pockets 
along $\Gamma-X$, should be shifted down by $\sim 23$ meV in the $B \parallel b$ direction and $\sim 46$ meV in the $B \parallel c$ direction.
The hole pocket along $S-Y$ should be shifted up by $\sim 2$ meV in the $B \parallel b$ direction. It is important to note that for a quantitiative 
agreement with ARPES and dHvA at $\theta=90^{\circ}$, this hole pocket should be pulled down such that it does not merge with the hole pocket along $\Gamma-X$.
We estimate this energy shift to be about $-5$ meV. Further, the open hole pocket along $Y-\Gamma$ should 
be shifted down by $\sim 19$ meV in both $B \parallel b$ and $B \parallel c$ direction such that it forms a needle-like closed FS, leading
to finite pocket size which would in turn lead to a finite dHvA frequency at $\theta=90^{\circ}$. 
As shown in Fig. S1, LDA with dense $k$-mesh also captures most of these features. However, 
quantitative agreemenet is poor. Details of required shifts for different cases is presented in the Supplementary Note III,\cite{esi} 
which suggests that GGA calculations for optimized structure best capture the electronic structure details, except for the open FS
along $\Gamma-Y$.

In comparison, the calculated dependence of dHvA frequencies on $\theta$ is much less messy in {\trt}, with the frequency range for 
hole pockets $H_1$ and $H_2$ of only 40T and 100T, respectively. These values are much smaller than in {\tit} due to the small size of the corresponding 
hole pockets along $\Gamma -X$. Interestingly, despite the presence of fish-and-toad hole pocket along $Y-\Gamma$, due to its smallness combined with
the absence of hole pocket along $S-Y$, we find a smooth curve for the needle shaped $H_3$. 
Similarly, the hole pocket $H_4$ along $\Gamma-Y$  has a substantially small FS area.
 
\subsection{Sensitivity of WPs to model approximations and implications for magnetotransport. }

Having established a qualitative and quantitative correspondence between the calculated and observed electronic structure details, we now discuss their 
contribution to
the observed magnetotransport. 
We begin with the following observations about the MR response in {\tit}:\cite{Khim2016} 
The reported quantum limit (QL) at $F_0 \approx 10$ T corresponds to a ``trivial" small fermion pocket which aligns with the observed phase intercept $\gamma=0$
in the LL fan diagram. 
At the same time, the DFT predicted contributions from the electron pockets close to $\Gamma$ 
(labeled $E_4$ and $E_5$ in Ref. \onlinecite{Khim2016}) were not observed.
This was ascribed to their large scattering rate(s). However, 
origin of such a large scattering rate is not clear. 
An anomalous non-saturating positive transverse MR $\propto B^{1.5}$ for fields upto 70 T was observed. 
We note that {\nit} also shows positive transverse MR which scales as $B^{1.25}$, however electron hole charge compensation is not the primary source
of large transverse MR based on two band model.\cite{Zhou2019,Schonemann2019}
Additionally, in the high field transverse MR data for $B \parallel b$ in {\tit}, between $50$ to $60$ T, a sharp plateau is observed, and beyond 
60 T, the MR jumps and shows a change in slope around 65 T. Unfortunately, clear explanations of these features are not available. 
At the same time, equivalent studies in {\nit} are also not available as the magnetic fields only up to 35 T were considered. 

The {\it tiered} WP landscape (Fig. \ref{fig:wp_ta}(a)) offers a unified interpretation of the anomalous transport in {\tit} in the following way:
Rather than treating the $10$ T Hall kink, the unexplained frequencies ($F_1, F_2$), missing electron pockets ($E_4, E_5$), 
and the non-saturating $B^{1.5}$ MR as independent anomalies, we suggest that they emerge collectively due to WP quartets proximate to  
the Fermi level, at $+9.6$ meV and $-46.3$ meV (GGA). 
The Quantum Limit at $\sim 10$ T with a sharp Hall kink is not just a geometric band-edge crossing, but is
concomitant to Berry curvature maximization driven by the Fermi energy Weyl nodes, at $\sim 10$ meV, effectively 'pinning' the carrier 
transport to the chiral $n=0$ LL. 
Considering $E_{\rm WP}= 10$ meV and a typical Fermi velocity, $v_F = 1.0\times 10^5$ m/s, we find that the magnetic 
field required to attain the 
quantum limit,\cite{Armitage2018, Lu2022} $B_{QL1} = E_{\rm WP}^2/2e\hbar v_{\rm F}^2 \approx 7.6$ T. Increasing to $v_F = 1.5\times 10^5$ m/s, reduces the required magnetic field  to $B_{QL1} = 3.4$ T. 

We further attribute the plateau in the MR at $50$ to $60$ T to the presence of $\sim 46$ meV WP quartet. The effective field required to 
reach the corresponding quantum limit is given by: 
$B_{QL2} \sim 72$T (using $v_F = 1.5 \times 10^5$ m/s).
Therefore, it is apparent that the predicted WP energies are in good agreement with the
observations considering the required resizing of the fermion pockets for a quantitative agreement.
The presence of multiple WPs may lead to field induced phase coherence between the 
inter-tier type-II WPs predicted here:
The observed sub-quadratic $B^{1.5}$ MR in {\tit} likely reflects a superposition of two competing
regimes driven by the tiered WP landscape. While the near-Fermi-level WP quartet ($P_1$ at $\sim 10$ meV)
enters its quantum limit at $\sim 7-11$ T, where single-LL
(Abrikosov-type) linear MR is expected,\cite{Abrikosov1998} the second quartet ($P_2$ at $\sim 46$ meV)
remains in the semiclassical regime over the same field range, contributing a $B^2$-like dependence.
The effective power law measured over the entire field up to 70 T can thus be an
intermediate exponent. A quantitative theory of this crossover, accounting for the specific node energies and
velocities, however, remains an open question.

The fact that a second set of WPs lie at $\sim -46$ meV, coincidental with the electron pockets in range $-20$ to $-50$ meV 
in the density functional calculations, further implies enhanced anomalous velocity 
which reduces the phase coherence of the corresponding (semiclassical) orbits, manifesting phenomenologically as the large effective masses and elevated 
scattering rates reported earlier.\cite{Khim2016}

Therefore, the (largest) shift by approximately $46$ meV ($P_2$) for the optimized structure, due to changes in atomic 
position (see Fig. \ref{fig:wp_ta} and Table S6) not withstanding, the energy positions of the Fermi energy WPs is in a 
good quantitative agreement with the reported anomalies in the MR, rendering the predicted WP landscape to be quite stable and robust. 
The observed $\gamma=0$ despite Fermi energy WPs predicted here can, in principle, arise due to Berry curvature cancellation from these two WPs of 
opposite chirality, $P_1$ at $+9.6$ meV with $\chi = -1$ and $P_2$ at $-46.3$ meV with $\chi = +1$ 
if the relevant cyclotron orbits enclose both these WPs, cancelling their nontrivial contributions.\cite{Arnold2016} 
At the same time, we note that possible phase cancellation from the Zeeman phase shift $\Delta \gamma \approx 0.5$ cannot be ruled out:\cite{Alexandradinata2017}
 Using the relation $\Delta \gamma = g^* m^*/4 m_0$ with $m^* \approx 0.1 m_0$, we obtain $g^* \approx 20$. 
 This large effective $g$-factor is consistent with the strong spin-orbit coupling inherent to systems involving heavy elements like Ir
 and Te. 

Combining the relatively sparse energy distribution of WP with the fact that the electron pockets 
 are strongly hybridized with $P_2$ leads to imbalance in the carrier distribution, overcoming the topological contribution of negative MR 
 (chiral anomaly) and resulting in the observed large positive MR in {\tit}. In comparison, in {\trt}, the emergence of the chiral anomaly 
is consistent with the prediction of a transition from a relatively sparse 
tiered WP landscape to a dense 
tier regime (many more WPs in the $\pm 30$ meV range). The reduction in inter-tier gaps leads to a breakdown of the phase-cancellation 
mechanism observed in {\tit}, potentially revealing a non-zero Berry phase intercept ($\gamma \neq 0$) in high-sensitivity dHvA measurements on {\trt}.

\subsection{Predictions for Nb-based compounds Nb$M$Te$_4$ ($M$=Ir, Rh). }

\begin{figure}[hbt]
    \centering
     \includegraphics[width=0.8\textwidth]{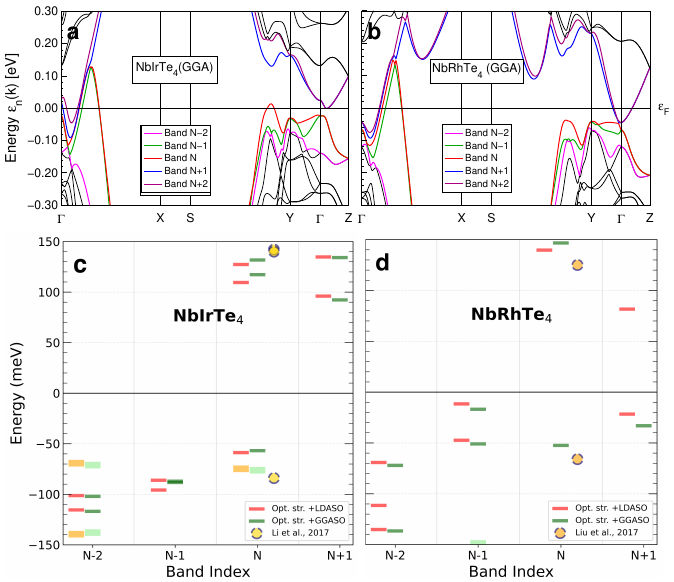}
	\caption{\small \textbf{Electronic structure and topology of Nb compounds.} \textbf{(a-b)} Electronic band structure for optimized crystal structure of (a) {\nit} and (b) {\nrt}. \textbf{(c-d)} WP landscape in (c) {\nit} and (d) {\nrt}. Other details same as in Fig. \ref{fig:wp_ta}.}
    \label{fig:wp_nit}
\end{figure}

To gain further insights into the topological electronic properties of the ternary tellurides, we further investigate the electronic properties
(bandstructure and Fermi surface topology) and the WP landscape of Nb-based compounds Nb$M$Te$_4$ ($M$=Ir, Rh). Figure \ref{fig:wp_nit}(a) shows the bandstructure 
for {\nit}. Compared to {\tit}, we note two distinct changes in the Fermi surface: First, the fish-and-toad 
structure is located at approximately $-25$ meV and are skewed such that the hole pockets along $\Gamma-Y$ are absent. 
Second, the size of the hole pocket along $S-Y$ are comparatively much smaller. The corresponding FS is shown in Fig. S12(a).
Concurrently, WPs are located atleast approximately $50$ meV away from the Fermi energy, as shown in Fig. \ref{fig:wp_nit}(c). 
Our results for energies of WPs from band $N$ compare very well with the earlier reports.\cite{Li2017} In total, we find 60 WPs (GGA)  
whose BZ positions are listed in Table S8.

The relatively smaller MR response in {\nit} compared to {\tit} ($\sim 80 \,\%$ vs $800\,\%$ for $\mathbf{B} \perp \mathbf{E}$) can be attributed to these 
distinct differences in their electronic structure. The absence of WPs within $\pm 50$ meV of the Fermi energy in {\nit} removes the Berry 
curvature enhanced transport and consequent amplification of MR in {\tit}. As the hole pocket along $S-Y$ is also present {\tit}, the absence of a proximate 
WP implies a trivial QL at low fields, rendering it difficult to observe. This is consistent with the absence of a 
clear quantum limit feature in measurements up to 35 T.\cite{Zhou2019} 

Further, the absence of the needle-like quasi-1D hole pocket along $\Gamma-Y$ in {\nit} removes a 
significant geometric contribution to the non-saturating MR that is present in {\tit}.\cite{Zhang2019} 
Furthermore, the large separation between WP tiers in {\nit} precludes the inter-tier contributions proposed here for {\tit}, leading to 
a modest  $B^{1.25}$ dependence and and a clean phase intercept in the LL fan diagram.\cite{Zhou2019}
These predictions can be directly tested: pockets hybridized with nearby WPs should exhibit anomalously large Dingle temperatures in quantum 
oscillation measurements, reflecting Berry curvature-enhanced scattering, while pockets far from any WP should show conventional behaviour. 
Measurements extending to higher fields in {\nit} would therefore provide a valuable benchmark for the role of WP proximity in magnetotransport.

The bandstructure for the sister compound {\nrt} is shown in Fig. \ref{fig:wp_nit}(b). 
Change of $M$ from Ir to Rh leads to similar qualitative changes as for the Ta compounds.
In comparison to {\trt}, the needle-like hole pocket along $\Gamma-Y$ is absent as the bands along this path of the BZ are shifted down 
compared to {\trt}. At the same time, the bands $N+1$ and $N+2$ are shifted higher along $S-Y$, leading to absence of electron pocket along this 
high symmetry path. The FS derived from these bands is shown in Fig. S12 (b). The obtained WP landscape is shown in Fig. \ref{fig:wp_nit}(d), and the positions 
of the WPs are listed in Table S9.
We further note that {\nrt} is a promising materials where WP driven transport could be observed as all the WPs are energetically well-separated.

\begin{figure}
    \centering
    \includegraphics[width=0.8\textwidth]{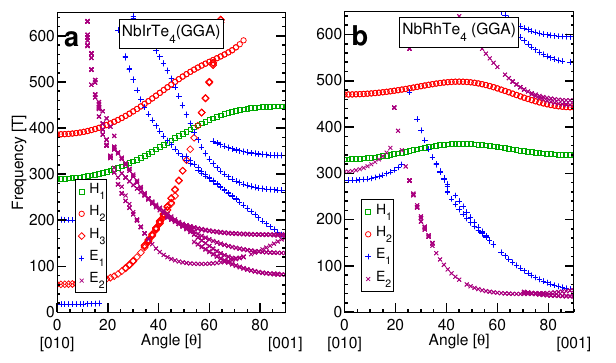}
	\caption{\small \textbf{Angular dependence of dHvA in Nb compounds.} \textbf{(a-b)} Calculated angular dependence of dHvA for optimized crystal structure of (a) {\nit} and (b) {\nrt}}
    \label{fig:dhva2}
\end{figure}

The Nb-based systems, therefore, provide a very useful contrast to the Ta-based systems. On one hand, the Fermi surfaces are significantly different. On the other hand, 
{\nrt} hosts a Fermi energy WP whereas {\nit} does not. The calculated angular dependence of dHvA frequencies is shown in Fig. \ref{fig:dhva2}, and is consistent with earlier report \cite{Zhou2019}.
In comparison with the Ta-based compounds, we observe the absence of curves $H_4$ and $H_5$ in both Ir and Rh systems due to the absence of 
no hole pocket along $Y-\Gamma$. 
Further, in {\nrt}, the curve corresponding to $H_3$ is also missing, since there is no hole pocket along $S-Y$. 
At the same, presence of electron pocket at $\Gamma$ due to band $N+1$ should be directly observable in experiments.
Signatures of Fermi energy WP in {\nrt} and absence of any small Fermi pocket could serve as a benchmark for the origin of 
the quantum limit in in presence of magnetic fields.

As far as the nature of WPs in the Nb compounds is concerned, the low-energy dispersions around the WPs from band $N$ in {\nit} reveals a 
rather unique situation, as shown in Fig. S9. 
It turns out that {\nit} hosts WPs of type-I, type-II as well as type-III. While the WP $W_1$ at $E_{W1} \sim 132$ meV is of type I with upright 
Weyl cone, $W_2$ and $W_3$, at energies $E_{W2} \sim 117$ meV and $E_{W3} \sim -57$ meV are type-II. 
The WP $W_4$ lies at $E_{W4} \sim -76$ meV, is 8-fold degenerate, and appears to be of type-III with critical tilt.\cite{Li2021}

At the same time, WPs from band $N$ in {\nrt} also hosts a WP quartet at $E_{W1} \sim 150$ meV with a critical tilt together with a type-II WP quartet
at $E_{W2} \sim -50$ meV, as shown in Fig. S10, rendering them as hybrid Weyl semimetals hosting WPs of different type. 
The other Fermi energy WPs from band $N\pm 1$ are type-II in nature, as shown in Fig. S11.
Therefore, Nb-compounds are interesting hybrid Weyl semimetals hosting WPs of different types, highlighting the complexity and richness of the ternary tellurides.

When compared with their Ta counterparts, we find that the nature of one of the WP quartet from band $N$ transforms from critically tilted in {\nrt} to type I 
when either or both transition metals ($A$ and $M$) are replaced by their partners with valence $5d$ shell (Nb $\rightarrow$ Ta, Rh $\rightarrow$ Ir). 
In other words, the nature of these WPs are tunable with increasing chemical pressure, opening up avenues for precise control of the WPs and their 
transport signatures. 

\section*{Outlook}
To summarize, by accounting for all the relevant sub-bands, we predict a rich WP landscape in $AM$Te$_4$ ($A$=Ta, Nb, $M$=Ir, Rh) compunds. 
The most noteworthy aspect is the presence of Fermi energy WPs, merely a few meV away from the Fermi energy, in all the considered 
compunds except {\nit}. These Fermi-energy WPs are well-separated and may lead to large observable Fermi arcs --- a feature directly measurable in ARPES. 
Our comparative study involving $k$-mesh dependence, structure optimization, DFT functional, and Wannier models further provides a much
needed map between the electronic structure, topology, and the available observations in these compounds. 

We find that GGA provides a much better description of the electronic structure in {\tit} where a very good agreement for the previously unexplained 
features of dHvA is found. We further quantified estimates for the the anisotropic shifts of the bands/fermion pockets 
for a quantitative agreement between DFT and experiments. Despite largest required shift of up to $\sim 40$ meV, the predicted energies and BZ positions of the WPs 
provides a consistent explanation of the observed quantum limit, absence of contributions from the electron pockets predicted in DFT, 
sub-quadratic growth and the observed $50$ to $60$ T plateau in MR data as arising from the ``tiered" WP landscape. 

The tiered WP landscape readily extends to other members of the family, whereby the observed chiral anomaly in {\trt} as well as the observed 
suppression of positive MR in {\nit} compared to {\tit} can also be explained on the basis of the WP landscape. Among all the considered 
compounds, we find that {\nrt} hosts a Fermi-energy WP which is energetically well-separated. 

It is equally interesting that all the considered compounds host WPs of different type, implying their hybrid nature. 
Both the Ta compounds (Ta$M$Te$_4$) host type-I and type-II WPs. Remarkably, WPs from band $N$ in Nb$M$Te$_4$ host critically tilted type-III WPs.
These critically tilted WPs transform to type-I when $4d$ transition metals are replaced by their $5d$ counterparts, suggesting tunability under chemical pressure.
These findings elucidate the rich and complex electronic structure topology of the ternary tellurides $AM$Te$_4$, positioning them as a fascinating materials class
to explore the fundamental aspects of Weyl points and their contributions to optical and transport responses.

The comparative study of ternary tellurides considered here constitutes an
important factor in analyzing their magnetoresponse in terms of the underlying topology of the electronic structure while
highlighting a common misconception in characterizing the topological electronic structure of Weyl semimetals in general. 
As such, this warrants a critical examination of the WP landscape in Weyl semimetals and their consequences, especially in di- and ternary-chalcogenides.
In addition, presence of Fermi-energy type-II WPs in a tiered structure may provide unique signatures in (linear and non-linear) 
transport and optical responses in these materials. Identifying them is important from a fundamental viewpoint which 
may also open up new avenues for harnessing the potential of WSMs for applications.

\section*{Methods}
We used the FPLO code\cite{Koepernik1999}, version 22.62 [\href{https://www.fplo.de}{https://www.fplo.de}], for 
DFT-based calculations with LDA\cite{PW1992} and GGA.\cite{Perdew1996} A $k$-mesh with $20\times10\times10$ intervals 
was used along with the tetrahedron method for the BZ integration. Full-relativistic calculations, accounting for spin-orbit 
coupling (SOC) effects, were performed via 4-spinor method as implemented in the FPLO code. To investigate the $k$-mesh 
dependence, we also used a $k$-mesh with $12 \times 6 \times 6$ intervals as in earlier DFT calculations of {\tit}. 
For Fermi surface and dHvA oscillations, a $k$-mesh with $120\times60\times60$ intervals was used.

For {\tit}, {\trt} and {\nit}, the external parameters (lattice constants) were kept fixed to their experimental values\cite{Koepernik2016, Shipunov2021, Lee2024}
while the atomic positions were optimized.  
For {\nrt}, the lattice constants of {\nit} were appropriately scaled to the same ratio as in the Ta compounds, and the atomic positions were optimized.
The residual forces in the relaxed structures were less than 1 meV/{\AA}.
A comparison of relaxed and experimental structures is tabulated in Supplementary Note I.\cite{esi}

To study the topological electronic structure, maximally projected Wannier fucntions (WFs) were used to construct a high-quality Wannier model 
which accurately captures the bandstructures of all the valence and conduction bands in the energy window $\sim -6.5$ eV to about 4 eV.
Difference between the resulting Wannier model and the self-consistent band structures was typically $\lesssim 2$, as shown in Fig. S1. 
The high quality Wannier model involves Ta($5d$), Ir($5d,6s$), Te($5p$) orbitals for {\tit}, Ta($5d$), Rh($4d,5s$), Te($5p$) orbitals for {\trt}, 
Nb($4d$), Ir($5d,6s$), Te($5p$) orbitals for {\nit}, and Nb($4d$), Rh($4d$), Te($5p$) orbitals for {\nrt}. This corresponds to 184 orbitals 
in the basis set for Wannier projections: For {\tit}, {\trt} and {\nit}, and 176 orbitals for {\nrt}. For Ta-based compounds a minimal wannier model was
also considered where the outermost $s$ orbitals of Ir/Rh were excluded. 
However, it turns out that the minimal model fails to accurately capture the fish-and-toad structure along $\Gamma-Y$ (see Fig. S1(g) \& (h)).

The WPs were obtained in the energy window of $-150.0$ meV to $+150.0$ meV using the PYFPLO interface of the FPLO code for different $k$-meshs.
They were subsequently verified by calculating the Chern number. The tilting of the Weyl cones and thus the WP type, especially for the Fermi energy WPs,
were ascertained by calculating the dispersions along the three principal axis centered at the WP, as detailed in Supplementary Note IV.\cite{esi}

\section*{Acknowledgements} 
We thank Ulrike Nitzche for maintaining the computational cluster at IFW Dresden, Germany, and Sanjay Kumar Prasad for maintaining the 
computational facility at BIT Mesra, where parts of the calculations were done. 
R.~R. and S.~P. also thank Anusandhan National Research Foundation (ANRF), Government of India, 
for development of the computational resources, vide Project code SRG/2022-23/001697, used in this study.
J.~v.~d.~B. Acknowledges support from the Deutsche Forschungsgemeinschaft (DFG, German Research Foundation) under Germany’s Excellence Strategy 
through the W\"urzburg-Dresden Cluster of Excellence on Complexity and Topology in Quantum Matter --- {\it ct.qmat} (EXC 2147, project-id 390858490). 
J.~v.~d.~B. also acknowledges support from the Deutsche Forschungsgemeinschaft (DFG, German Research Foundation) within the Collaborative Research 
Center ``Correlated Magnetism: From Frustration to Topology'' (SFB 1143, project-id 247310070).

\section*{Author contributions}
J.~v.~d.~B. and R.~R. conceptualized the work and developed the work plan. S.~P. carried out the calculations with help from R.~R.. 
S.~P. and R.~R. prepared the first version of the manuscript, J.~v.~d.~B. revised the manuscript. All authors discussed and analyzed the results.

\section*{Data Availability}
The data that support the findings of this study are available from the corresponding
author upon reasonable request.

\section*{Competing interests} 
The authors declare no competing interests.

\section*{Correspondence} 
All correspondence related to the manuscript should be made to \href{j.van.den.brink@ifw-dresden.de}{j.van.den.brink@ifw-dresden.de} (J.~v.~d.~B.) and \href{r.ray@bitmesra.ac.in}{r.ray@bitmesra.ac.in} (R.~R.).

\vspace{2cm}

\clearpage
\newpage
\begin{center}
    \textbf{Supplementary Information}
\end{center}

%
%
%
%
%

\newcolumntype{Y}{>{\centering\arraybackslash}X}
\newcolumntype{w}{>{\centering\arraybackslash}p{0.90cm}}	
\newcolumntype{n}{>{\centering\arraybackslash}p{0.25cm}}	



\renewcommand{\thesection}{Suppementary Note \Roman{section}:}
\renewcommand{\thetable}{S\arabic{table}}
\renewcommand{\thefigure}{S\arabic{figure}}

\section{Structural Details}

The DFT calculations were carried out for $AM$Te$_4$ ($A$=Ta, Nb, $M$=Ir, Rh).
For Ir-based {\tit} and {\nit}, the reported experimental structures were used.\cite{Koepernik2016,Lee2024} In addition, the atomic positions were optimized using local 
density approximation (LDA)\cite{PW1992} and generalized gradient approximation (GGA)\cite{Perdew1996} while keeping the lattice constants fixed to the experimental values.
The force threshold of 0.001 ev/\AA was used. For {\trt}, the reported structure\cite{Shipunov2021} possesss a large residual forces, upto 0.07 eV/\AA. 
As a result, only a structure with the same lattice parameters as the experimental structure, and optimized atomic positions were used.
The lattice parameters of {\nrt} were estimated by scaling the experimental lattice constants of {\nit} by the same ratios as that of 
the lattice constants of the experimental structures of {\tit} \& {\trt}. The atomic positions were relaxed while keeping the lattice constants fixed. 
The crystal structure details include lattice parameters, Wyckoff positions, corresponding atoms and their 
fractional coordinates ($x, y, z$) and are listed in Tables \ref{tab:str_tit}, \ref{tab:str_trt} \& \ref{tab:str_nmt}. 

For {\tit} the largest deviation in atomic positions between experimental and LDA-optimized structure is observed to be 0.042 {\AA}, and 0.046 {\AA} between 
the LDA- and GGA-optimized structures. Whereas, for {\trt}, the maximum shift between experimental and LDA-optimized structure is 0.388 {\AA}. 
Between, LDA and GGA-optimized structures, only a difference of 0.046 {\AA} was found. 

\section{Parameter Dependence}

\begin{figure}
\centering
\includegraphics[width=0.95\textwidth]{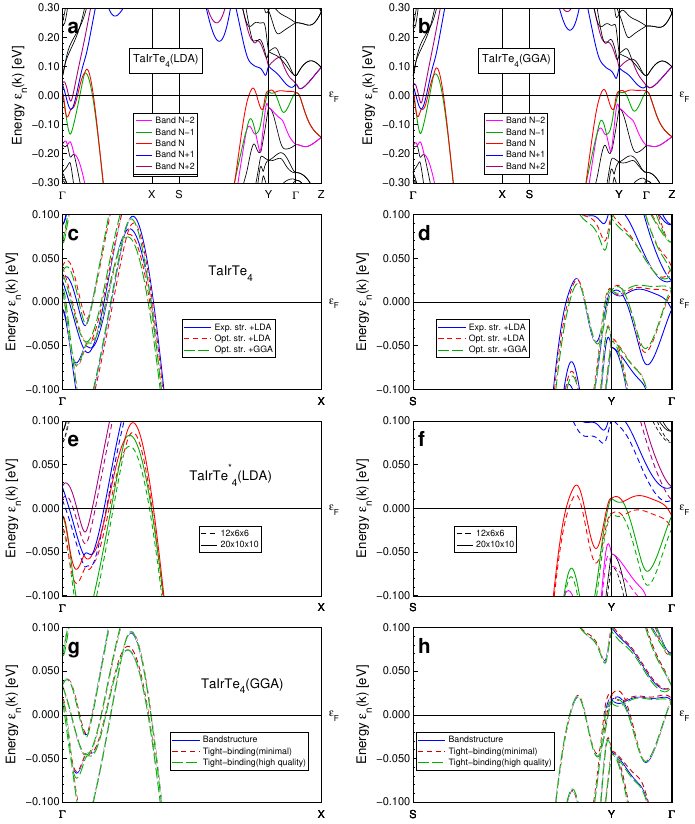}
	\caption{A detailed comparison of the electronic properties (bandstructures) of {\tit} between different crystal structures, $k$-mesh, 
	and XC functional. Note that the optimized structures (Opt str.) correspond to 
	LDA-optimized structures. Spin-orbit effects were included.}
\label{fig:si_elprop_tit}
\end{figure}

\begin{figure}
\centering
\includegraphics[width=0.99\textwidth]{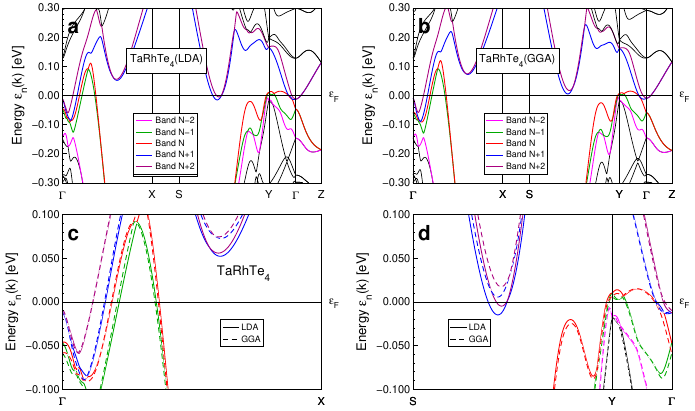}
	\caption{A detailed comparison of the electronic properties (bandstructure) of {\trt} between different $k$-mesh and XC functionals for the LDA-optimized structures of {\trt}.
	Spin-orbit effects were included.}
\label{fig:si_elprop_trt}
\end{figure}

In Figs. \ref{fig:si_elprop_tit} and \ref{fig:si_elprop_trt}, we show the sensititvity of the calculated bandstructures to various DFT 
parameters for {\tit} and {\trt}. Using {\tit} we provide a detailed comparison of the electronic properties (bandstructure) between experimental 
and LDA-optimised crystal structures. Additionally we also show the effects of different $k$-mesh and XC functional in our study. 
For {\tit}, we also provide a comparison between different Wannier models --- minimal basis vs full valence orbital basis. 
The minimal basis Wannier model was constructed by excluding the $(n+1) s$ orbital of Ir.  

\section{Quantum Oscillations (dHvA)}

\begin{figure}
\centering
\includegraphics[width=0.85\textwidth]{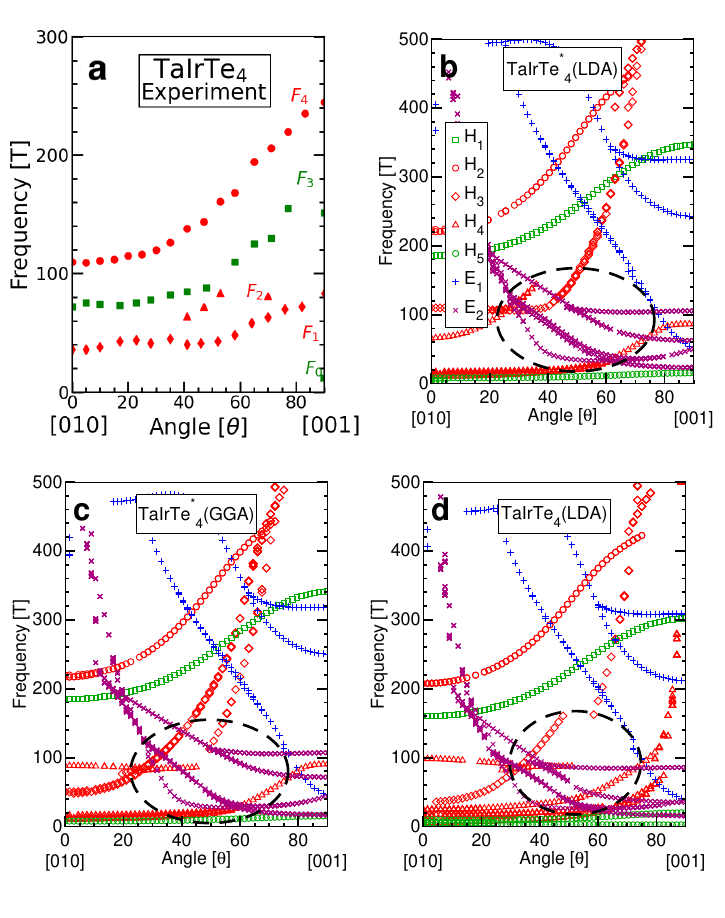}
	\caption{A detailed comparison of the dHvA frequencies between observed different crystal structures and XC functional. The experimental data is shown in (a). (b) and (c) correspond to the 
	DFT calculations using the experimental crystal structure with a dense $k$-mesh, while (d) corresponds to DFT calculations on the LDA-optimized structure using LDA.}
\label{fig:si_dhva_tit}
\end{figure}

In Figs. \ref{fig:si_dhva_tit}, we show and compare the de Haas-van Alphen (dHvA) frequencies of {\tit} obtained from the 
experiments (Fig. \ref{fig:si_dhva_tit} (a)),\cite{Khim2016}  with our calculations on experimental structure using a $k$-mesh 
with $120 \times 60 \times 60$ intervals along with LDA (Fig. \ref{fig:si_dhva_tit} (b)) and GGA (Fig. \ref{fig:si_dhva_tit} (c)) fuctionals.
The curves $H_1$ corresponds to $F_3$, $H_2$ to $F_4$, $H_3$ to $F_2$, and $H_4$ to $F_1$. Since, the hole-pockets $H_2$ and $H_3$ along $\Gamma-X$
and $S-Y$, merges in our calculations, the dHvA frequencies shoots up along $B \parallel c$ direction. Interestingly, in the experimental structure
the needle-like hole-pocket along $Y-\Gamma$ does not have an open Fermi surface, therefore, the corresponding dHvA frequency for $H_4$ does not shoot 
up and gives a finite value ($\sim 90$ T) along $B \parallel c$ direction.

In  Fig. \ref{fig:si_dhva_tit} (d), we also show the dHvA frequencies for the optimized crystal structure of {\tit} using a $k$-mesh 
with $120 \times 60 \times 60$ intervals and LDA fuctional. The behaviour of dHvA frequencies are similar to the results obtained 
through GGA on an optimized structure using a $k$-mesh with $20 \times 10 \times 10$ intervals. The curves $H_2$ and $H_3$ correspoding 
to hole-pockets along $\Gamma-X$ and $S-Y$, sees a sharp increase along $B \parallel c$ direction due to merging of pockets. The curve
$H_4$ also shoots up, as the needle-like hole-pocket along $Y-\Gamma$ has an open Fermi surface in the optimized structures. 

Further, as shown in Table \ref{tab:dhva_tit}, we evaluated the energies associated with these frequencies using the Onsager relation.\cite{Onsager1952} 
In Table \ref{tab:dhva_calc_tit}, we present the energy shifts requried in the calculated hole pockets to have a good agreement with the experimental observations. 

\section{Weyl Point Characteristics \& Type}

In Table \ref{tab:wp_tit}, \ref{tab:wp_trt}, \ref{tab:wp_nit} \& \ref{tab:wp_nrt}, we provide a complete description of Weyl points (WPs) 
identified in {\tit}, {\trt}, {\nit} and {\nrt}. For each WP, we specify the band contributions, energy values, their chiralities, Brillouin 
zone positions, and their degeneracy. 

Additionally, for selected WPs close to the Fermi energy ($|E_{WP}| < 50$ meV), we determine the WP-type. For this purpose, we study the dispersion along the 
principle directions. 
In other words, for a WP located at ($k_x, k_y, k_z$), we obtained the WP dispersion in these three directions: 
\begin{itemize}
\item Line passing through WP, $\parallel$ to $x$-axis : ($0, k_y, 0$) $\rightarrow$ ($k_x, k_y, k_z$) $\rightarrow$ ($1/2, k_y, 0$)
\item Line passing through WP, $\parallel$ to $y$-axis : ($k_x, 0, 0$) $\rightarrow$ ($k_x, k_y, k_z$) $\rightarrow$ ($k_x, 1/2, 0$)
\item Line passing through WP, $\parallel$ to $z$-axis : ($k_x, k_y, 1/4$) $\rightarrow$ ($k_x, k_y, k_z$) $\rightarrow$ ($1/2, k_y, -1/4$)
\end{itemize}

\begin{figure}
\centering
\includegraphics[width=\textwidth]{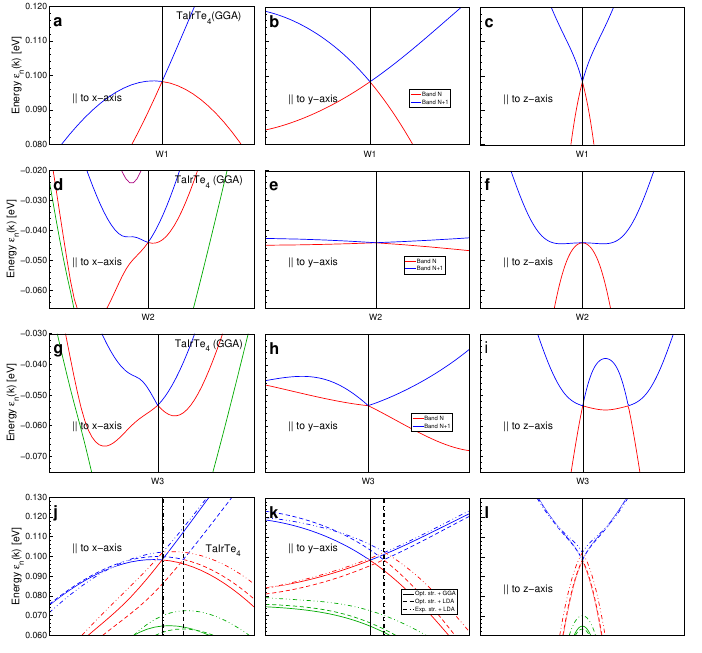}
	\caption{Low-energy dispersions along the three principal directions for the WPs from band (a-l) $N$ in {\tit}, $W_1$ - $W_3$. All the three WPs are found to be of Type-II.
	The BZ positions of these WPs is listed in Table \ref{tab:wp_tit}. Panels (j)-(l) show the comparison between the nature of WP $W_1$ between different structures, and XC functional, showing their evolution from type-II to type-I for the optimized structure.}
\label{fig:si_wptype_tit1}
\end{figure}

\begin{figure}
\centering
\includegraphics[width=\textwidth]{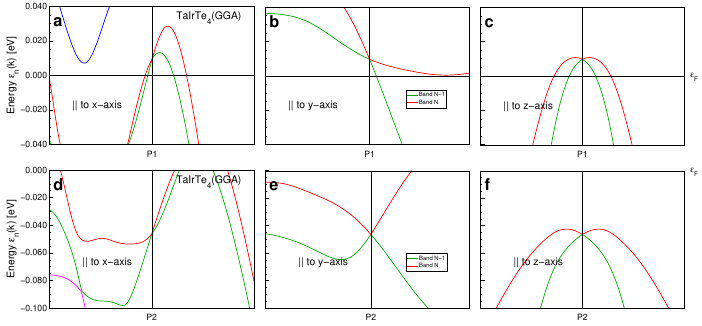}
	\caption{Low-energy dispersions along the three principal directions for the Fermi energy WPs from band (a-f) $N-1$ in {\tit}, $P_1$ and $P_2$. Both the WPs are found to be of type-II. 
	The BZ positions of these WPs is listed in Table \ref{tab:wp_tit}.}
\label{fig:si_wptype_tit2}
\end{figure}

\begin{figure}
\centering
\includegraphics[width=\textwidth]{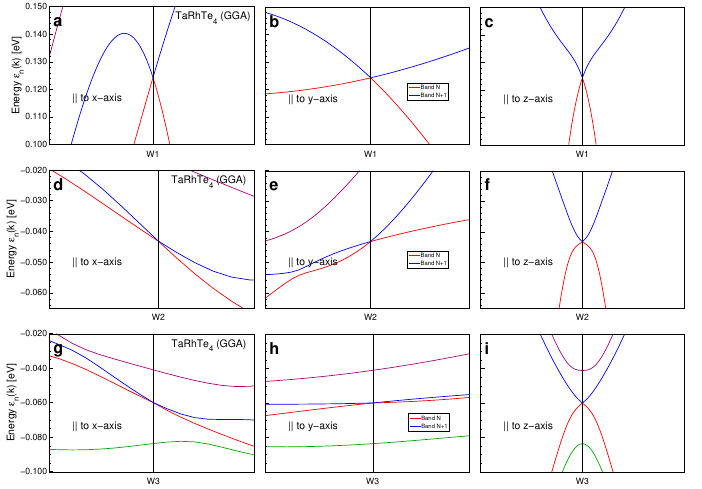}
	\caption{Low-energy dispersions along the three principal directions for the WPs from band (a-i) $N$ in {\trt}, $W_1$ - $W_3$. $W_1$ is a type-I WP while $W_2$ and $W_3$ are of type-II.  
	The BZ positions of these WPs is listed in Table \ref{tab:wp_trt}.}
\label{fig:si_wptype_trt1}
\end{figure}

\begin{figure}
\centering
\includegraphics[width=\textwidth]{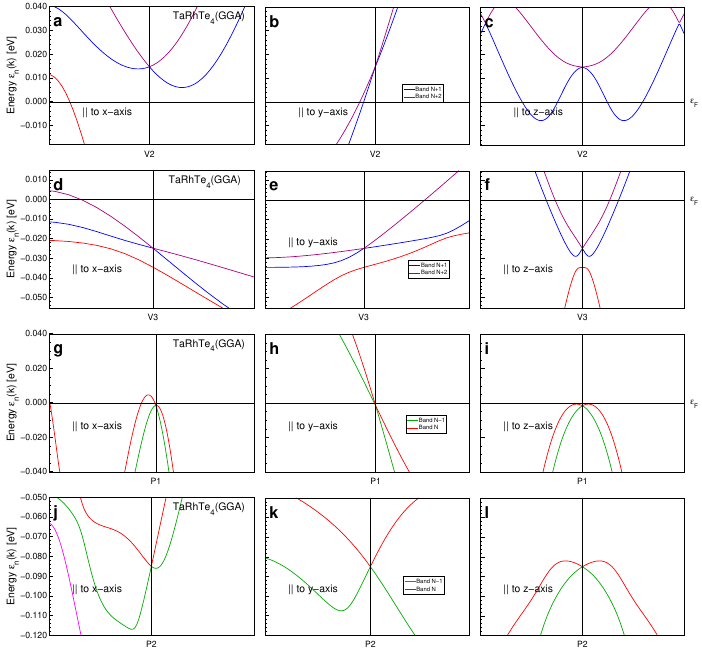}
	\caption{Low-energy dispersions along the three principal directions for the WPs from band (a-f) $N+1$ and (g-l) $N-1$ in {\trt}, $V_2$, $V_3$, $P_1$ and $P_2$. All the four WP are of type-II.  
	The BZ positions of these WPs is listed in Table \ref{tab:wp_trt}.}
\label{fig:si_wptype_trt2}
\end{figure}

\begin{figure}
\centering
\includegraphics[width=\textwidth]{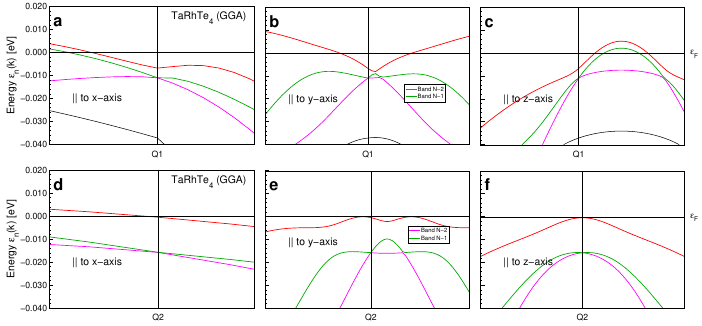}
	\caption{Low-energy dispersions along the three principal directions for the WPs from band (a-f) $N-2$ in {\trt}, $Q_1$ and $Q_2$. Both WPs are of type-II.  
	The BZ positions of these WPs is listed in Table \ref{tab:wp_trt}.}
\label{fig:si_wptype_trt3}
\end{figure}

\begin{figure}
\centering
\includegraphics[width=\textwidth]{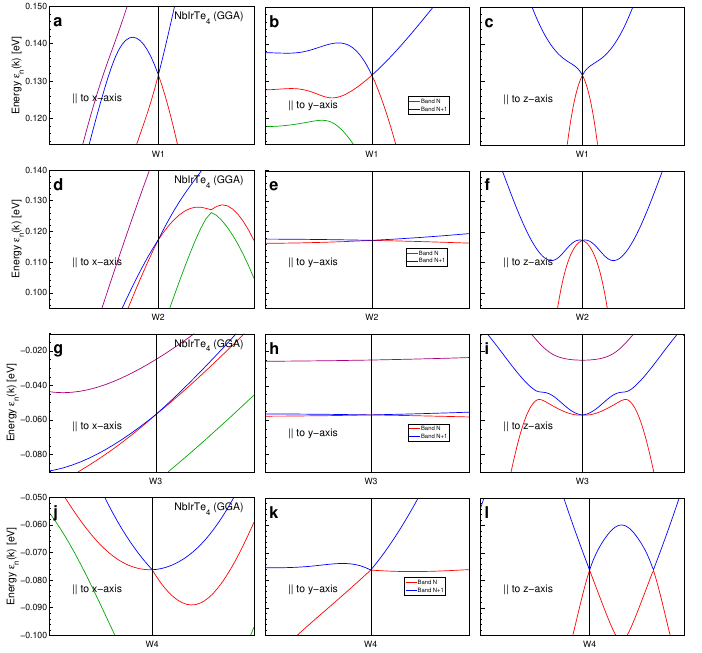}
	\caption{Low-energy dispersions along the three principal directions for the WPs from band (a-l) $N$ in {\nit}, $W_1$ - $W_4$. $W_1$ is a type-I WP while $W_2$ and $W_3$ are of type-II and $W_4$ appears to be of type-III. 
	The BZ positions of these WPs is listed in Table \ref{tab:wp_nit}.}
\label{fig:si_wptype_nit}
\end{figure}

\begin{figure}
\centering
\includegraphics[width=\textwidth]{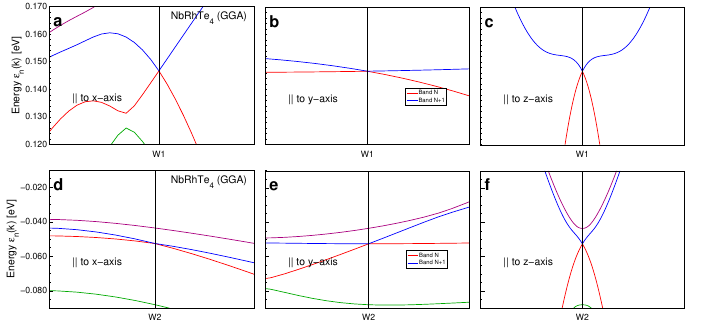}
	\caption{Low-energy dispersions along the three principal directions for the WPs from band (a-f) $N$ in {\nrt}, $W_1$ and $W_2$. $W_1$ is a type-I WP while $W_2$ is of type-II.  
	The BZ positions of these WPs is listed in Table \ref{tab:wp_nrt}.}
\label{fig:si_wptype_nrt1}
\end{figure}

\begin{figure}
\centering
\includegraphics[width=\textwidth]{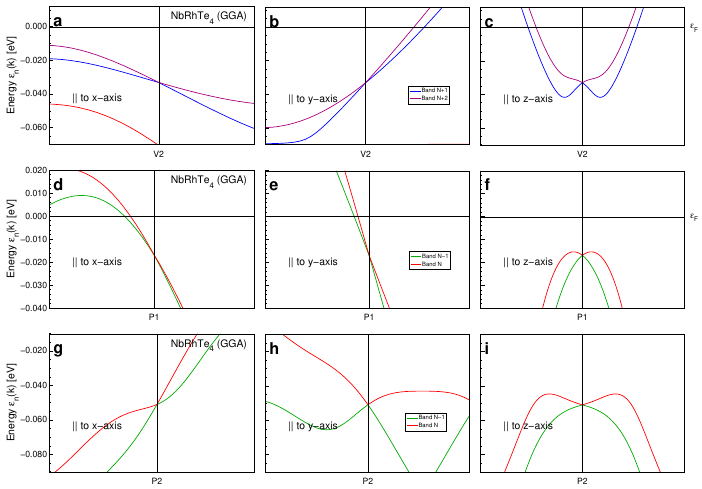}
	\caption{Low-energy dispersions along the three principal directions for the WPs from band (a-c) $N+1$ and (d-i) $N-1$ in {\nrt}, $V_2$, $P_1$ and $P_2$. All the three WPs are of type-II.  
	The BZ positions of these WPs is listed in Table \ref{tab:wp_nrt}.}
\label{fig:si_wptype_nrt2}
\end{figure}

Figures \ref{fig:si_wptype_tit1} - \ref{fig:si_wptype_nrt2} show the low-energy dispersion of these WPs. 
Note that in the figures, dispersions are shown only along a segment of the above lines.

\section{Fermi Surface of Nb-Compounds}

\begin{figure}
\centering
\includegraphics[width=\textwidth]{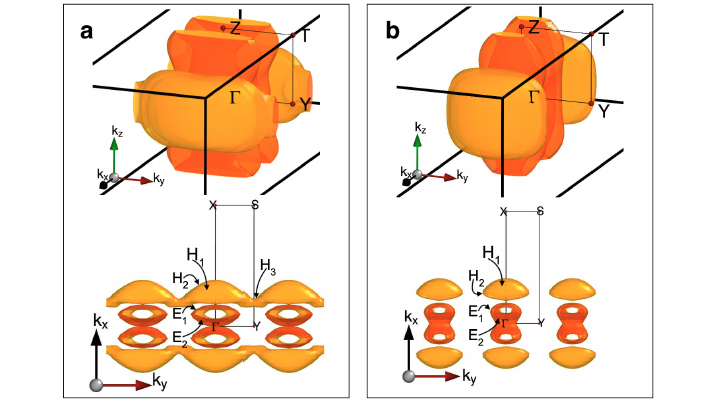}
	\caption{Calculated Fermi surface of Nb$M$Te$_4$ ($M$=Ir, Rh) within GGA, (a) {\nit} and (b) {\nrt}. We used a $k$-mesh with $120 \times 60 \times 60$ intervals to calculate the FS.}
\label{fig:si_fs_nb}
\end{figure}

In Fig. \ref{fig:si_fs_nb}, we show the calculated FS of Nb$M$Te$_4$ ($M$= Ir, Rh) using a $k$-mesh with $120 \times 60 \times 60$ intervals together with
GGA functional. Similar to its Ta counterpart, {\nit} also has merging of hole-pockets $H_2$ and $H_3$. As a result of this merging, we observe a sharp increase in
the dHvA frequencies of these pockets along $B \parallel c$ direction. Since the fish-and-toad structure is $\sim 25$ meV below the Fermi level, a needle-like dispersion 
is not observed here.

{\nrt} shares similar features in the FS as {\trt}, with the exception of few missing hole pockets. The small electron-pocket along $S-Y$ and the needle-like 
dispersion along $Y-\Gamma$  are absent here, because the bands responsible for these structures are away from the Fermi level in {\nrt} and do not 
contribute to the FS. 

\begin{table}
\centering
\normalfont
        \caption{\small Structural details of {\tit}. Fractional atomic coordinates ($x,y,z$) in the experimental, 
	LDA-optimized and GGA-optimized structures. The lattice constants are kept fixed: $a= 3.77$ {\AA}, $b= 12.421$ {\AA}, $c= 13.184$ {\AA}.}
\label{tab:str_tit}
	\renewcommand{\arraystretch}{1.50}
	\small
	\begin{tabularx}{1.0\textwidth}{p{0.75cm} p{1.05cm} p{0.5cm} p{1.20cm} p{1.20 cm} p{0.25cm} p{0.5cm} p{1.20cm} p{1.20cm} p{0.25cm} p{0.5cm} p{1.20cm} p{1.20cm}}
\hline
\hline
		& & \multicolumn{3}{c}{Experimental str. } & & \multicolumn{3}{c}{LDA-Opt str.} & & \multicolumn{3}{c}{GGA-Opt str.}  \\
		\cline{3-5} 
		\cline{7-9} 
		\cline{11-13} 
		\textbf{No.} & \textbf{Atom} & $x$ & $y$ & $z$ & & $x$ & $y$ & $z$ & & $x$ & $y$ & $z$ \\
\hline
\hline
		1 & Ta & 0 & 0.0540 & 0.0041 & & 0 & 0.0545 & 0.0041 & & 0 & 0.0548 & 0.0041 \\
		2 & Ta & 0 & 0.2697 & 0.4910 & & 0 & 0.2713 & 0.4901 & & 0 & 0.2707 & 0.4900\\
		3 & Ir &0 & 0.5355 & 0 & & 0 & 0.5359 & 0.9989 & & 0 & 0.5367 & 0.9991\\
		4 & Ir & 0 & 0.7543 & 0.4916 & & 0 & 0.7570 & 0.4903 & & 0 & 0.7577 & 0.4896\\
		5 & Te & 0 & 0.0648 & 0.3897 & & 0 & 0.0662 & 0.3905 & & 0 & 0.0658 & 0.3878\\
		6 & Te & 0 & 0.1937 & 0.8527 & & 0 & 0.1923 & 0.8500 & & 0 & 0.1919 & 0.8465\\
		7 & Te & 0 & 0.3458 & 0.0957 & & 0 &  0.3443 & 0.0950 & & 0 & 0.3445 & 0.0966\\
		8 & Te  & 0 & 0.4142 & 0.6391 & & 0 & 0.4148 & 0.6392 & & 0 & 0.4142 & 0.6423\\
		9 & Te  & 0 & 0.5642 & 0.3960 & & 0 & 0.5652 & 0.3936 & & 0 & 0.5652 & 0.3914\\
		10 & Te & 0 & 0.6776 & 0.8469 & & 0 & 0.6782 & 0.8447 & & 0 & 0.6787 & 0.8420\\
		11 & Te & 0 & 0.8492 & 0.1078 & & 0 & 0.8487 & 0.1062 & & 0 & 0.8493 & 0.1088\\
		12 & Te & 0 & 0.8933 & 0.6484 & & 0 & 0.8951 & 0.6495 & & 0 & 0.8954 & 0.6522\\
\hline
\end{tabularx}
\end{table}

\begin{table}
\centering
\normalfont
        \caption{\small Structural details of {\trt}. Fractional atomic coordinates ($x,y,z$) in the experimental, LDA-optimized and GGA-optimized structures, with the 
	lattice constants: $a= 3.757$ {\AA}, $b= 12.548$ {\AA}, $c= 13.166$ {\AA}.}
\label{tab:str_trt}
	\renewcommand{\arraystretch}{1.50}
\small
	\begin{tabularx}{1.0\textwidth}{p{0.75cm} p{1.05cm} p{0.5cm} p{1.20cm} p{1.20 cm} p{0.25cm} p{0.5cm} p{1.20cm} p{1.20cm} p{0.25cm} p{0.5cm} p{1.20cm} p{1.20cm}}
\hline
\hline
		& & \multicolumn{3}{c}{Experimental str. } & & \multicolumn{3}{c}{LDA-Opt str.} & & \multicolumn{3}{c}{GGA-Opt str.}  \\
		\cline{3-5} 
		\cline{7-9} 
		\cline{11-13} 
		\textbf{No.} & \textbf{Atom} & $x$ & $y$ & $z$ & & $x$ & $y$ & $z$ & & $x$ & $y$ & $z$ \\
\hline
\hline
		1 & Ta & 0 & 0.0558 & 0.8080 & & 0 & 0.0531 &  0.8080 & & 0 & 0.0534 & 0.8080  \\
		2 & Ta & 0 & 0.2653 & 0.2910  & & 0 & 0.2708 & 0.2950 & & 0 & 0.2701 & 0.2951\\
		3 & Rh &0 & 0.5276 & 0.8020 & & 0 & 0.5347 & 0.8033 & & 0 & 0.5354 & 0.8036\\
		4 & Rh & 0 & 0.7543 & 0.3310 & & 0 & 0.7559 & 0.2950 & & 0 & 0.7563 & 0.2944\\
		5 & Te & 0 & 0.0595 & 0.1905 & & 0 & 0.0675 & 0.1954 & & 0 & 0.0671 & 0.1927\\
		6 & Te & 0 & 0.1945 & 0.6260 & & 0 & 0.1914 &  0.6555 & & 0 & 0.1908 & 0.6520\\
		7 & Te & 0 & 0.3448 & 0.9081 & & 0 & 0.3452 & 0.8980 & & 0 & 0.3457 & 0.9001\\
		8 & Te  & 0 & 0.4078 & 0.4448 & & 0 & 0.4139 &  0.4431 & & 0 & 0.4129 & 0.4466\\
		9 & Te  & 0 & 0.5727 & 0.1998 & & 0 & 0.5666 & 0.1996 & & 0 & 0.5665 & 0.1971\\
		10 & Te & 0 & 0.6773 & 0.6576 & & 0 & 0.6761 &  0.6502 & & 0 & 0.6769 & 0.6475\\
		11 & Te & 0 & 0.8504 & 0.9226 & & 0 & 0.8492 & 0.9104 & & 0 & 0.8499 & 0.9131\\
		12 & Te & 0 & 0.9044 & 0.4675 & & 0 & 0.8937 & 0.4525 & & 0 & 0.8942 & 0.4553\\
\hline
\end{tabularx}
\end{table}

\begin{table}
\centering
\normalfont
        \caption{\small Structural details of Nb$M$Te$_4$ (M= Ir, Rh). Fractional atomic coordinates ($x,y,z$) of 
			LDA-optimized structures, with the lattice constants: $a= 3.77$ {\AA}, $b= 12.51$ {\AA}, $c= 13.12$ {\AA} for {\nit},  
			and $a= 3.757$ {\AA}, $b= 12.638$ \AA, $c= 13.102$ {\AA} for {\nrt}.}
\label{tab:str_nmt}
	\renewcommand{\arraystretch}{1.50}
\small
	\begin{tabularx}{1.0\textwidth}{p{1.5cm} p {1.9cm} p{1.0cm} p{1.5cm} p{1.0cm} p{1.0cm} p{1.5cm}  p{1.5cm} p{1.5cm}}
\hline
\hline
		& & \multicolumn{3}{c}{{\nit}} & & \multicolumn{3}{c}{{\nrt}}  \\
		\cline{3-5}
		\cline{7-9}
		\textbf{No.} & \textbf{Atom} & $x$ & $y$ & $z$ & & $x$ & $y$ & $z$ \\
\hline
\hline
		1 & Nb & 0 & 0.0544 & 0.0041 & & 0 & 0.0536 & -0.1920 \\
		2 & Nb & 0 & 0.2719 & 0.4900 & & 0 & 0.2717 & 0.2948 \\
		3 & Ir/Rh & 0 & 0.5650 & 0.9985 & & 0 & 0.5339 & 0.8028 \\
		4 & Ir/Rh & 0 & 0.7561 & 0.4903 & & 0 & 0.7550 & 0.2948 \\
		5 & Te & 0 & 0.0676 & 0.3903 & & 0 & 0.0688 & 0.1955 \\
		6 & Te & 0 & 0.1924 & 0.8499 & & 0 & 0.1917 & 0.6550 \\
		7 & Te & 0 & 0.3445 & 0.0951 & & 0 & 0.3454 & 0.8980 \\
		8 & Te & 0 & 0.4157 & 0.6390 & & 0 & 0.4149 & 0.4429 \\
		9 & Te & 0 & 0.5655 & 0.3929 & & 0 & 0.5668 & 0.1988 \\
		10 & Te & 0 & 0.6762 & 0.8438 & & 0 & 0.6741 & 0.6489 \\
                11 & Te & 0 & 0.8494 & 0.1069 & & 0 & 0.8499 & 0.9107 \\
		12 & Te & 0 & 0.8929 & 0.6500 & & 0 & 0.8914 & 0.4533 \\
\hline
\end{tabularx}
\end{table}

\begin{table}
\centering
\normalfont
        \caption{\small Comparison of the Fermi velocity ($v_{\rm F}$), dHvA frequencies and their associated energy values for different magnetic field orientations, extracted from the observed dHvA data and DFT simulations. 
	Observed frequencies (curves) are labeled $F_i$, while the electron and hole pockets are labeled $E_i$ and $H_i$, repectively.}
\label{tab:dhva_tit}
	\renewcommand{\arraystretch}{1.50}
\small
	\begin{tabularx}{\textwidth}{p{2.5cm} p{1.2cm} p{1.5cm} p{1.5cm} p{0.25cm} p{1.5cm} p{1.0cm} p{0.25cm} p{1.5cm} p{1.0cm}}
\hline
\hline
		& \textbf{Curve} & \multicolumn{2}{c}{\textbf{Fermi Velocity (m/s)}} & & \multicolumn{2}{c}{\textbf{Frequency (T)}} & & \multicolumn{2}{c}{\textbf{Energy (meV)}} \\
		\cline{3-4}
		\cline{6-7}
		\cline{9-10}

		&  & \textbf{B$\parallel$b} & \textbf{B$\parallel$c} & &  \textbf{B$\parallel$b} & \textbf{B$\parallel$c} & & \textbf{B$\parallel$b} & \textbf{B$\parallel$c}\\
\hline
\hline
		& F1 & $1.92 * 10^5$ & $2.87 * 10^5$ & & 36 & 82 & & 40.2 & 90.6 \\
		\textbf{Experimental} & F2 & $1.04 * 10^5$ &               & & 36 &    & & 21.8 &  \\
		\textbf{Data} & F3 & $1.48 * 10^5$ & $2.12 * 10^5$ & & 73.5 & 150 & & 44.3 & 90.6 \\
		& F4 & $1.80 * 10^5$ & $2.67 * 10^5$ & & 109 & 245 & & 65.5 & 144.3 \\
 \hline
		& H1 & $1.48 * 10^5$ & $2.12 * 10^5$ & & 185 & 345 & & 70.2 & 137.3 \\
		\textbf{Exp. str.}& H2 & $1.80 * 10^5$ & $2.67 * 10^5$ & & 210 & 460 & & 91.0 & 199.7 \\
		\textbf{(LDA)}& H3 & $1.04 * 10^5$ &               & & 110 &     & & 38 &  \\
		& H4 & $1.92 * 10^5$ & $2.87 * 10^5$ & & 65 & 85 & & 54.0 & 92.3 \\
 \hline
		& H1 & $1.48 * 10^5$ & $2.12 * 10^5$ & & 185 & 340 & & 70.2 & 136.3 \\
		\textbf{Exp. str.} & H2 & $1.80 * 10^5$ & $2.67 * 10^5$ & & 220 & 440 & & 93.1 & 195.3 \\
		\textbf{(GGA)} & H3 & $1.04 * 10^5$ &               & & 50 &     & & 25.6 &  \\
		& H4 & $1.92 * 10^5$ & $2.87 * 10^5$ & & 90 & 90 & & 63.5 & 95.0 \\
 \hline
		& H1 & $1.48 * 10^5$ & $2.12 * 10^5$ & & 160 & 310 & & 65.3 & 130.2 \\
		\textbf{Opt. str.} & H2 & $1.80 * 10^5$ & $2.67 * 10^5$ & & 210 & 430 & & 91.0 & 193.1 \\
		\textbf{(LDA)} & H3 & $1.04 * 10^5$ &              & & 40 &      & & 22.9 &  \\
		& H4 & $1.92 * 10^5$ & $2.87 * 10^5$ & & 100 & 140 & & 67.0 & 118.4 \\
 \hline
		& H1 & $1.48 * 10^5$ & $2.12 * 10^5$ & & 160 & 300 & & 65.3 & 128.1 \\
		\textbf{Opt. str.} & H2 & $1.80 * 10^5$ & $2.67 * 10^5$ & & 200 & 420 & & 88.8 & 190.8 \\
		\textbf{(GGA)} & H3 & $1.04 * 10^5$ &              & & 30 &      & & 19.9 &  \\
		& H4 & $1.92 * 10^5$ & $2.87 * 10^5$ & & 75 & 120 & & 58.0 & 109.6 \\
 \hline
\end{tabularx}
\end{table}

\begin{table}
\centering
\normalfont
	\caption{\small Calculated energy shifts, $\delta E = E_{\rm Exp} - E_{\rm DFT}$, between the observed dHvA frequencies ($F_i$) and the calculated 
	frequencies from various fermion pockets ($H_i$ and $E_i$), required to match our DFT results with experimental values. Extracted parameters from Table \ref{tab:dhva_tit} were used.}
\label{tab:dhva_calc_tit}
	\renewcommand{\arraystretch}{1.50}
\small
	\begin{tabularx}{\textwidth}{p{1.42cm} p{1.2cm} p{1.2cm} p{0.20cm} p{1.2cm} p{1.2cm} p{0.20cm} p{1.2cm} p{1.2cm} p{0.20cm} p{1.2cm} p{1.2cm}}
\hline\hline
 \textbf{Curve} & \multicolumn{11}{c}{\textbf{Energy Difference (meV)}}\\ 
		\cline{2-12}
		& \multicolumn{2}{c}{\textbf{Exp.-str. (LDA)}} & & \multicolumn{2}{c}{\textbf{Exp.-str. (GGA)}} & & \multicolumn{2}{c}{\textbf{Opt.-str. (LDA)}} & & \multicolumn{2}{c}{\textbf{Opt.-str. (GGA)}} \\
		\cline{2-3} \cline{5-6}\cline{8-9}\cline{11-12}
		& \textbf{B$\parallel$b} & \textbf{B$\parallel$c} & & \textbf{B$\parallel$b} & \textbf{B$\parallel$c} & & \textbf{B$\parallel$b} & \textbf{B$\parallel$c} & & \textbf{B$\parallel$b} & \textbf{B$\parallel$c}\\
		\hline\hline
		F3-H1 & -26.0 & -46.8 & & -26.0 & -45.8  & & -21.0 & -39.6 & & -21.0 & -37.5\\
		F4-H2 & -25.4 & -55.5 & & -27.6 & -51.1 & & -25.4 & -48.8 & & -23.2 & -46.6\\
		F2-H3 & -16.3 &       & & -3.9 &        & & -1.2  &       & & 1.9   &      \\
		F1-H4 & -13.8 & -1.6 & & -23.3 & -4.3   & & -26.8 & -27.8 & & -17.8 & -19.0\\
 \hline
\end{tabularx}
\end{table}

\clearpage
\newpage

\begin{sidewaystable}
\centering
        \caption{\small WP landscape in {\tit}. Band, Energy (E), BZ position, chirality ($\chi$) and type of WP in {\tit}. Comparison is drawn between the results of LDA calculation for the experimental(Exp.-str.) and optimized structure (Opt.-str.) and GGA calculation for the optimized structure. The energies $E$ are in meV, the BZ positions are in the units of ${2\pi}/a$, and the degeneracy of the WPs is denoted by $d$. Spin orbit effects were included for all cases.}
\label{tab:wp_tit}
	\renewcommand{\arraystretch}{1.50}
\small
        \begin{tabularx}{1\textwidth}{p{0.6cm} p {0.50cm} p{0.50cm} p{1.0cm} p{0.7cm} p{0.7cm} p{0.7cm} p{0.4cm} p{0.15cm} p{1.0cm} p{0.7cm}  p{0.7cm} p{0.7cm} p{0.4cm} p{0.15cm} p{1.0cm} p{0.7cm} p{0.7cm} p{0.7cm} p{0.4cm} p{0.65cm}}
\hline \hline
                & & & \multicolumn{5}{c}{\textbf{Exp.-str. (LDA)}} & & \multicolumn{5}{c}{\textbf{Opt.-str. (LDA)}} & &\multicolumn{5}{c}{\textbf{Opt.-str. (GGA)}} & \\ \cline{4-8} \cline{10-14} \cline{16-20}
                \textbf{Band} & \textbf{WP} & $\chi$ & $E$ & $k_x$ & $k_y$ & $k_z$ & $d$ & & $E'$ & $k_x'$ & $k_y'$ & $k_z'$ & $d'$ & & $E"$ & $k_x"$ & $k_y"$ & $k_z"$ & $d"$ & \textbf{Type}  \\
\hline \hline
                N   & W$_1$ & -1 & 102.3  & 0.122  & 0.168 & 0.000 & 4 & & 98.5 & 0.126 & 0.051 & 0.000 & 4 & & 98.4 & 0.122 & 0.046 & 0.000 & 4 & I \\ 
                N   & W$_2$ & -1 & & & & & & & & & & & & & -44.1 & 0.054 & 0.008 & 0.000 & 4 & II \\ 
                N   & W$_3$ &  1 & & & & & & & & & & & & & -53.3 & 0.047 & 0.017 & 0.034 & 8 & II \\ 
                N-1 & P$_1$ & -1 & 7.2    & 0.104 & 0.101 & 0.000 & 4 & & 12.6 & 0.110 & 0.095 & 0.000 & 4 & & 9.6 & 0.106 & 0.092 & 0.000 & 4 & II \\
                N-1 & P$_2$ &  1 & -0.2   & 0.101 & 0.102 & 0.000 & 4 & & -37.2 & 0.084 & 0.094 & 0.000 & 4 & & -46.3 & 0.080 & 0.090 & 0.000 & 4 & II \\ 
                N-1 & P$_3$ &  1 & -84.2  & 0.062 & 0.058 & 0.000 & 4 & & -89.0 & 0.059 & 0.063 & 0.000 & 4 & & -93.1 & 0.058 & 0.067 & 0.000 & 4 & \\ 
                N-1 & P$_4$ & -1 & -104.1 & 0.055 & 0.058 & 0.063 & 8 & & -87.4 & 0.029 & 0.056 &  0.059 & 8 & & -82.4 & 0.027 & 0.052 & 0.055 & 8 &  \\ 
                N-1 & P$_5$ &  1 & -110.3 & 0.039 & 0.050 & 0.070 & 8 & & -102.8 &  0.057 & 0.060 & 0.063 & 8 & & -100.2 & 0.058 & 0.061 & 0.057 & 8 & \\ 
                N-2 & Q$_1$ &  1 & -98.1  & 0.019 & 0.067 & 0.000 & 4 & & -81.4 & 0.017 & 0.072 & 0.000 & 4 & & -81.5 & 0.019 & 0.075 & 0.000 & 4 & \\ 
                N-2 & Q$_2$ &  1 & -106.5 & 0.028 & 0.100 & 0.039 & 8 & & -91.1 & 0.028 & 0.090 & 0.000 & 4 & & -86.6 & 0.026 & 0.092 & 0.000 & 4 & \\ 
                N-2 & Q$_3$ & -1 & -107.2 & 0.010 & 0.083 & 0.054 & 8 & & -97.6 & 0.064 & 0.070 & 0.000 & 4 & & -98.21 & 0.060 & 0.071 & 0.000 & 4 & \\ 
                N-2 & Q$_4$ & -1 & -112.9 & 0.031 & 0.083 & 0.000 & 4 & & -100.8 & 0.059 & 0.071 & 0.012 & 8 & & -99.6 & 0.059 & 0.072 & 0.005 & 8 & \\ 
                N-2 & Q$_5$ & -1 & -134.9 & 0.043 & 0.039 & 0.000 & 4 & & -130.5 & 0.045 & 0.036 & 0.000 & 4 & & -131.2 & 0.045 & 0.039 & 0.000 & 4 & \\  
                N-2 & Q$_6$ & -1 & -148.0 & 0.016 & 0.069 & 0.110 & 8 & & -142.0 & 0.007 & 0.077 & 0.124 & 8 & & & & & &  \\ 
\hline
\end{tabularx}
\end{sidewaystable}

\begin{table}
\centering
\normalfont
	\caption{\small WP landscape in {\trt}. Band, Energy (E), BZ position, chirality ($\chi$) and type of WP in {\trt}. Comparison is drawn between the results of LDA and GGA calculation for the optimized structure (Opt.-str.). 
	 The energies $E$ are in meV, the BZ position are in the units of $2\pi/a$, and degeneracy of the WPs is denoted by $d$. Spin-orbit effects were included for all cases.}
\label{tab:wp_trt}
\small
		\begin{tabularx}{1.0\textwidth}{p{0.6cm} p{0.33cm} p{0.33cm} p{0.004cm} p{1cm} w w w n p{0.004cm} p{1cm} w w w n n }
\hline\hline
			& & & & \multicolumn{5}{c}{\bf Opt-str. (LDA)} & & \multicolumn{5}{c}{\bf Opt-str. (GGA)}   \\
		\cline{5-9}
		\cline{11-15}
		\textbf{Band} & \textbf{WP} & $\chi$  & & $E$ & $k_x$ & $k_y$ & $k_z$ & $d$ & & $E'$ & $k_x'$ & $k_y'$ & $k_z'$ & $d'$ & \textbf{Type}\\
\hline \hline 
		N+1 & V$_1$ & -1 & & 25.3 & 0.274 & 0.079 & 0.000 & 4 & & 52.4 & 0.269 & 0.077 & 0.000 & 4 & \\
		N+1 & V$_2$ & 1 & & 14.3 & 0.043 & 0.064 & 0.000 & 4 & & 14.7 & 0.040 & 0.067 & 0.000 & 4 &  II \\
		N+1 & V$_3$ & 1 & & -17.5 & 0.010 & 0.020 & 0.000 & 4 & & -24.8 & 0.013 & 0.022 & 0.000 & 4 &  II \\
		N & W$_1$ & -1 & & 116.0 & 0.154 & 0.038 & 0.000 & 4 & & 124.4 & 0.151 & 0.033 & 0.000 & 4 &  I \\
		N & W$_2$ & 1 & & -44.2 & 0.017 & 0.016 & 0.000 & 4 & & -43.1 & 0.020 & 0.033 & 0.000 & 4 &  II \\
		N & W$_3$ & 1 & & -63.1 & 0.026 & 0.023 & 0.000 & 4 & & -59.9 & 0.022 & 0.018 & 0.000 & 4 &  II \\
		N-1 & P$_1$ & -1 & & -2.6 & 0.137 & 0.092 & 0.000 & 4 & & -1.5 & 0.135 & 0.091 & 0.000 & 4 &  II \\
		N-1 & P$_2$ & 1 & &-79.4 & 0.082 & 0.083 & 0.000 & 4 & & -85.0  & 0.080 & 0.081 & 0.000 & 4 &  II \\
		N-1 & P$_3$ & 1 & & & & & & & & -91.4 & 0.032 & 0.029 & 0.026 & 8 &  \\
		N-1 & P$_4$ & -1 & & & & & & & & -114.3 & 0.035 & 0.035 & 0.048 & 8 &  \\
		N-1 & P$_5$ & -1 & & -133.4 & 0.060 & 0.059 & 0.071 & 8 & & -131.9 & 0.062 & 0.059 & 0.067 & 8 & \\
		N-1 & P$_6$ & 1 & & -137.7 & 0.046 & 0.046 & 0.072 & 8 & & -133.1 & 0.041 & 0.041 & 0.064 & 8 & \\
			N-2 & Q$_1$ & -1 & & -12.5 & 0.005 & 0.145 & 0.023 & 8 & & -11.0 & 0.005 & 0.147 & 0.021 & 8 & II \\
		N-2 & Q$_2$ & 1 & & -19.0 & 0.0136 & 0.136 & 0.000 & 4 & & -15.7 & 0.012 & 0.140 & 0.000 & 4 & II  \\
		N-2 & Q$_3$ & 1 & & -114.7 & 0.069 & 0.061 & 0.000 & 4 & & -116.8  & 0.067 & 0.063  & 0.000 & 4 & \\
\hline
\end{tabularx}
\end{table}
 
\begin{table}
\centering
\normalfont
        \caption{\small WP landscape in {\nit}. Band, Energy (E), BZ position, chirality ($\chi$) and type of WP in {\nit}. Comparison is drawn between the results of LDA and GGA calculation for the optimized structure (Opt.-str.). The energies $E$ are in meV, the BZ positions are in the units of ${2\pi}/a$, and the degeneracy of the WPs is denoted by $d$. Spin orbit effects were included for all cases.}
\label{tab:wp_nit}
\small
		\begin{tabularx}{1.0\textwidth}{p{0.6cm} p{0.33cm} p{0.33cm} p{0.004cm} p{1cm} w w w n p{0.004cm} p{1cm} w w w n n }
\hline\hline
			& & & & \multicolumn{5}{c}{\bf Opt-str. (LDA)} & & \multicolumn{5}{c}{\bf Opt-str. (GGA)}   \\
		\cline{5-9}
		\cline{11-15}

                \textbf{Band} & \textbf{WP} & $\chi$  & & $E$ & $k_x$ & $k_y$ & $k_z$ & $d$ & & $E'$ & $k_x'$ & $k_y'$ & $k_z'$ & $d'$ & \textbf{Type}\\
\hline \hline
N+1 & V$_1$ & -1 & & 134.7 & 0.127 & 0.054 & 0.000 & 4 & & 134.1 & 0.126 & 0.048 & 0.000 & 4 & \\
N+1 & V$_2$ & 1 & & 96.1 & 0.112 & 0.050 & 0.000 & 4 & & 92.2 & 0.111 & 0.042 & 0.000 & 4 & \\
N & W$_1$ & -1 & & 127.3 & 0.147 & 0.055 & 0.000 & 4 & & 131.7 & 0.145 & 0.052 & 0.000 & 4 & I\\
N & W$_2$ & -1 & & 109.5 & 0.134 & 0.013 & 0.000 & 4 & & 117.2 & 0.134 & 0.008 & 0.000 & 4 & II\\
N & W$_3$ & -1 & & -58.8 & 0.072 & 0.009 & 0.000 & 4 & & -56.9 & 0.071 & 0.006 & 0.000 & 4 & II\\
N & W$_4$ & 1 & & -74.8 & 0.034 & 0.023 & 0.048 & 8 & & -76.2 & 0.034 & 0.024 & 0.047 & 8 & III\\
N-1 & P$_1$ & 1 & & -86.2 & 0.024 & 0.056 & 0.000 & 4 & & -88.4 & 0.025 & 0.057  & 0.000 &  4 & \\
N-1 & P$_2$ & 1 & & -95.8 & 0.200 & 0.052 & 0.000 & 4 & & -87.1 & 0.201 & 0.040 & 0.000 & 4 & \\
N-2 & Q$_1$ & -1 & & -69.2 & 0.013 & 0.141 & 0.036 & 8 & & -71.2 & 0.014 & 0.140 & 0.036 & 8 & \\
N-2 & Q$_2$ & 1 & & -101.3 & 0.028 & 0.106 & 0.000 & 4 & & -102.1 & 0.027 & 0.107 & 0.000 & 4 & \\
N-2 & Q$_3$ & 1 & & -115.5 & 0.065 & 0.044 & 0.000 & 4 & & -116.9 & 0.066 & 0.044 & 0.000 & 4 & \\
N-2 & Q$_4$ & 1 & & -139.4 & 0.042 & 0.075 & 0.057 & 8 & & -137.9 & 0.042 & 0.070 & 0.051 & 8 & \\
\hline

\end{tabularx}
\end{table}

\begin{table}
\centering
\normalfont
        \caption{\small WP landscape in {\nrt}. Band, Energy (E), BZ position, chirality ($\chi$) and type of WP in {\nrt}. Comparison is drawn between the results of LDA and GGA calculation for the optimized structure (Opt.-str.). The energies $E$ are in meV, the BZ positons are in the units of ${2\pi}/a$, and the  degeneracy of the WPs is denoted by $d$. Spin orbit effects were included for all cases.}
\label{tab:wp_nrt}
\small
		\begin{tabularx}{1.0\textwidth}{p{0.6cm} p{0.33cm} p{0.33cm} p{0.004cm} p{1cm} w w w n p{0.004cm} p{1cm} w w w n n }
\hline\hline
			& & & & \multicolumn{5}{c}{\bf Opt-str. (LDA)} & & \multicolumn{5}{c}{\bf Opt-str. (GGA)}   \\
		\cline{5-9}
		\cline{11-15}
                \textbf{Band} & \textbf{WP} & $\chi$  & & $E$ & $k_x$ & $k_y$ & $k_z$ & $d$ & & $E'$ & $k_x'$ & $k_y'$ & $k_z'$ & $d'$ & \textbf{Type}\\
\hline \hline
N+1 & V$_1$ & 1 & & 81.6 & 0.038 & 0.080 & 0.000 & 4 & & & & & & & \\
N+1 & V$_2$ & -1 & & -21.7 & 0.026 & 0.035 & 0.000 & 4 & & -33.1 & 0.021 & 0.030 & 0.000 & 4 & II\\
N & W$_1$ & -1 & & 139.6 & 0.173 & 0.024 & 0.000 & 4 & & 146.7 & 0.171 & 0.020 & 0.000 & 4 & III\\
N & W$_2$ & 1 & & & & & & & & -52.5 & 0.010 & 0.013 & 0.000 & 4 & II\\
N-1 & P$_1$ & -1 & & -11.6 & 0.160 & 0.098 & 0.000 & 4 & & -16.9 & 0.163 & 0.096 & 0.000 & 4 & II\\
N-1 & P$_2$ & 1 & & -47.4 & 0.105 & 0.102 & 0.000 & 4 & & -51.0 & 0.103 & 0.101 & 0.000 & 4 & II\\
N-1 & P$_3$ & -1 & & & & & & & & -148.9 & 0.005 & 0.006 & 0.058 & 8 &\\
N-2 & Q$_1$ & 1 & & -69.3 & 0.020 & 0.135 & 0.000 & 4 & & -72.1 & 0.020 & 0.136 & 0.000 & 4 &\\
N-2 & Q$_2$ & -1 & & -111.5 & 0.025 & 0.007 & 0.000 & 4 & & & & & & &\\
N-2 & Q$_3$ & 1 & & -135.2 & 0.077 & 0.042 & 0.000 & 4 & & -136.6 & 0.076 & 0.043 & 0.000 & 4 &\\
\hline

\end{tabularx}
\end{table}

\clearpage
\newpage


\end{document}